\documentclass{emulateapj}
\usepackage[usenames]{color}
\usepackage{epstopdf}

\defcitealias{AFG08}{Paper I}

\shorttitle{Detailed polarimetric properties of the Pipe Nebula}
\shortauthors{G.~A.~P. Franco et al.}

\begin{document}

\title{Detailed interstellar polarimetric properties of the Pipe nebula at core 
scales\slugcomment{Based on observations collected at Observat\'orio do Pico dos 
Dias, operated by Laborat\'orio Nacional de Astrof\'\i sica (LNA/MCT, 
Brazil).}}

\author{G.~A.~P. Franco}
\affil{Departamento de F\'\i sica -- ICEx -- UFMG, Caixa Postal 702, 30.123-970
Belo Horizonte, Brazil}
\email{franco@fisica.ufmg.br}

\author{F.~O. Alves and J.~M. Girart}
\affil{Institut de Ci\`encies de l'Espai (CSIC -- IEEC), Campus UAB--Facultat de
Ci\`encies, Torre C5--Parell 2$^\mathrm{a}$, 08193 Bellaterra, Catalunya, 
Spain}
\email{[oliveira;girart]@ieec.uab.es}


\begin{abstract}
We use $R$-band CCD linear polarimetry collected for about 12\,000 background
field stars in 46 fields of view toward the Pipe nebula to 
investigate the properties of the polarization across this dark cloud. Based on 
archival 2MASS data we estimate that the surveyed areas present total visual 
extinctions in the range $0\fm6 \le A_V \le 4\fm6$. While the observed 
polarizations show a well ordered large scale pattern, with polarization 
vectors almost perpendicularly aligned to the cloud's long axis, at core scales 
one see details that are characteristics of each core. Although many observed 
stars present degree of polarization which are unusual for the common interstellar 
medium, our analysis suggests that the dust grains constituting the diffuse 
parts of the Pipe nebula seem to have the same properties as the normal 
Galactic interstellar medium. Estimates of the second--order structure 
function of the polarization angles suggest that most of the Pipe nebula 
is magnetically dominated and that turbulence is sub-Alv\'enic. 
The Pipe nebula is certainly an interesting region where to investigate 
the processes prevailing during the initial phases of low mass stellar formation.
\end{abstract}

\keywords{ISM: clouds  --- ISM: individual objects: Pipe nebula  --- Stars: formation
--- Techniques: polarimetry}

\section{Introduction}\label{int}

The relatively low Galactic star formation efficiency (SFE, defined
as the fraction of a molecular gas mass that is converted into stars) 
is one fundamental constraint on the global properties of star formation.
In our Galaxy the SFE is observationally estimated to be of the order
of a few percent when whole giant molecular complexes are considered. For 
instance, the detailed study of the Taurus molecular cloud
complex conducted by \citet{GHN08} provided a SFE between 0.3 and 1.2\%. 
Magnetic fields and supersonic turbulence are two mechanisms that are commonly 
invoked for regulation of such small SFE. Magnetic fields may 
regulate cloud fragmentation by several physical processes (e.g., 
moderating the  infalling motions on the density peaks, controlling angular 
momentum evolution  through magnetic breaking, launching jets from the 
near-protostellar environment, etc). On the other hand, it is known that
turbulence may play a dual role, both creating overdensities to initiate 
gravitational contraction or collapse, and countering the effects of gravity
in these overdense regions. The respective rules of  magnetic fields and 
interstellar turbulence in regulating the core/star formation process are, 
however, highly controversial. For instance, some authors 
opine that magnetic fields are absolutely dominant in the star
formation process \citep[e.g.,][]{TM05, GLS06}, while others support that
super-Alfv\'enic turbulence provides a good description of molecular cloud
dynamics, and that the average magnetic field strength in those clouds
may be much smaller than required to support them against the gravitational
collapse \citep[see][and references therein]{PJJ04}.

The Pipe nebula, a massive filamentary cloud complex 
\citep[$10^4$ M$\sun$,][]{LAL06} located at the solar vicinity 
\citep[145\,pc,][]{AF07} which presents an apparently quiescent nature, 
seems to be an interesting place to look for some answers on the 
physical processes involved in the collapse of dense cloud cores and 
how they evolve until stars are formed. 

\begin{figure*}
\plotone{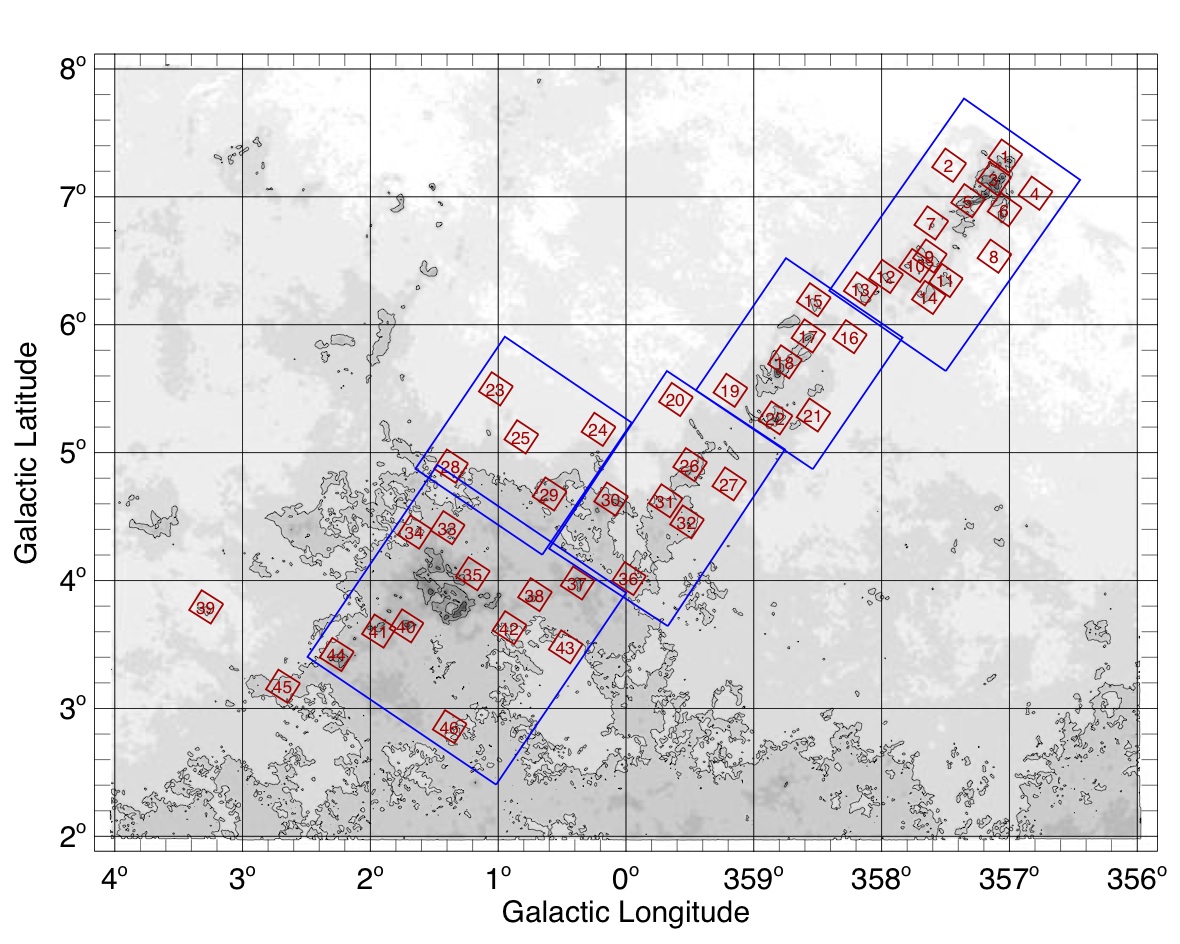}
\caption{Identification of the observed 46 lines-of-sight overplotted on
the dust extinction map of the Pipe nebula obtained by \citet{LAL06}. The small
squares roughly indicates the observed CCD field of view, which in our case
corresponds to about $12^\prime \times 12^\prime$. The large retangles 
demarcate the areas detailed separately in Figs.\,\ref{field_a} to \ref{field_d} 
(colored version of this and of the other figures are available in the online 
version of this paper).}
\label{pipe_areas}
\end{figure*}

Optical images of the Pipe nebula (see for instance the wonderful high quality 
image obtained by St\'ephane Guisard for the GigaGalaxy 
project\footnote{\tt http://www.gigagalaxyzoom.org})  or the dust extinction map 
obtained by  \citet{LAL06}, show that this complex comprises many dark cores 
and sinuous dark lanes. Although \citet{ALL07} have identified 159 dense 
cores with estimated masses ranging from 0.5 to 28 M$\sun$ all over 
the entire Pipe nebula, the only known star forming active site in 
this nebula seems to be restricted to its northwestern 
extreme (in galactic coordinates), the densest part of the complex associated 
with the dark cloud Barnard 59 (B\,59), which corresponds to only a small 
fraction of the entire cloud mass. Actually, an embedded cluster of young stellar
objects within B\,59 has been revealed by infrared images obtained with
the {\it Spitzer Space Telescope} \citep{BHB07}. The apparently low 
efficiency in forming stars observed for this cloud complex \citep[only $\sim$0.06\%
according to][]{FLM09,FPC10}, suggests that the Pipe nebula is an example of 
a molecular cloud in a very early stage of star formation. Indeed, in our previous 
paper \citep[][hereafter Paper I]{AFG08} it has been suggested that the Pipe 
nebula may present three distinct evolutionary stages, being the B\,59 region
the most evolved of them while the opposite extreme of the cloud (the
{\it bowl}) would be in the earlier stage. This suggestion seems to be reinforced by
the recent {\it Spitzer} census of star formation activity performed by \citet{FLM09},
who detected only six candidate young stellar objects (YSOs) outside the B\,59 
region, four of them located in the ``stem'' of the Pipe, none having been detected 
in the {\it bowl}. Moreover, the youthfulness of the YSOs in B\,59 is corroborated by
the results obtained by \citet{CLR10}, who estimated a median age of about
2.6\,Myrs to the candidate YSOs found in B\,59. Interestingly, they suggest that 
this population may be older than the well studied ones in Chamaeleon, Taurus, 
and $\rho$ Ophiuchus, respectively.

In \citetalias{AFG08} we described the global polarimetric properties of the Pipe 
nebula as obtained from mean values of polarization degree and dispersion in 
polarization angles calculated for stars having ${\rm P}/
\sigma_P \ge 10$. In the present paper we introduce the details of our data 
sample collected for 46 CCD fields, which are exactly the 
same as the one used in the previous work, and analyse the polarimetric 
properties of  the Pipe nebula at core scales. In order to 
increase the statistical sample for each investigated field, we were less strict in our 
selection criteria accepting stars with ${\rm P}/\sigma_P \ge 5$. 

\break

\section{Observations}\label{obs}

\subsection{Data acquisition and reductions}

\begin{deluxetable*}{crccc}
\tablecaption{Observed zero polarization standard stars.\label{table:1}}
\tablewidth{0pt}
\tablehead{
 &  &  \citeauthor{TBW90} & \citeauthor{SEL92} & this work \\
\multicolumn{1}{c}{HD} & \multicolumn{1}{c}{V} & P$_{V}$ (\%) & P$_{V}$ (\%) & P$_{R}$ (\%)}
\startdata
   12021 & 8.85 & --- & 0.078\,(0.018) & 0.106\,(0.037) \\
   98161 & 6.27 & 0.017\,(0.006) & --- & 0.028\,(0.041)\\
 154892 & 8.00 & 0.050\,(0.030) & --- & 0.027\,(0.041) \\
 176425 & 6.23 & 0.020\,(0.009) & --- & 0.031\,(0.017) \\
 BD+28\,\,4211 & 10.51 & --- & 0.054\,(0.027) & 0.066\,(0.025)
\enddata
\end{deluxetable*}

\begin{deluxetable*}{ccrrrrrr}
\tablecaption{Observed high polarization standard stars.\label{table:2}}
\tablewidth{0pt}
\tablehead{
 &  &  \multicolumn{2}{c}{\citeauthor{TBW90}} & \multicolumn{2}{c}{\citeauthor{heiles}} & \multicolumn{2}{c}{this work} \\
HD & V & \multicolumn{1}{c}{P$_{V}$ (\%)} & \multicolumn{1}{c}{$\theta$ (\degr)} & \multicolumn{1}{c}{P (\%)} & \multicolumn{1}{c}{$\theta$ (\degr)} & \multicolumn{1}{c}{P$_{R}$ (\%)} & \multicolumn{1}{c}{$\theta$ (\degr)}}
\startdata
 110984 & 8.95 & 5.70\,(0.01)& 91.6 & 5.19\,(0.11) & 90.6 & 5.21\,(0.19) & 91.4 \\
 111579 & 9.50 & 6.46\,(0.01)& 103.1 & 6.21\,(0.17) & 103.0 &  6.11\,(0.09) & 103.1 \\
 126593 & 8.50 & 5.02\,(0.01)& 75.2 & 4.27\,(0.10) & 77.0 & 4.65\,(0.11) & 74.9 \\
 155197 & 9.20 & 4.38\,(0.03)& 103.2 & 3.99\,(0.08) & 103.9 & 3.98\,(0.07) & 105.2 \\
 161306 & 8.30 &  \multicolumn{2}{c}{---}  & 3.69\,(0.09) & 67.5 & 3.59\,(0.23) & 67.9 \\
 168625 & 8.40 &  \multicolumn{2}{c}{---}  & 4.42\,(0.20) & 14.0 & 4.23\,(0.07) & 14.9 \\
 170938 & 7.90 &  \multicolumn{2}{c}{---}  & 3.69\,(0.20) & 119.0 & 3.62\,(0.11) & 118.8 \\
 172252 & 9.50 &  \multicolumn{2}{c}{---}  & 4.65\,(0.20) & 148.0 & 4.38\,(0.14) & 147.7 
\enddata
\end{deluxetable*}

The polarimetric data were collected with the 1.6\,m and the IAG 60\,cm 
telescopes at Observat\'orio do Pico dos Dias (LNA/MCT, Brazil) in missions 
conducted from 2005 to 2007. These data were obtained with the use of a 
specially adapted CCD camera to allow polarimetric measurements --- for a 
suitable description of the polarimeter see \citet{AM96}. $R$-band linear 
polarimetry was obtained for 46 fields (with field of view of about 
$12' \times 12'$ each) distributed over more than 7\degr\ (17\,pc in 
projection) covering the main body of the Pipe nebula. The
observing lines-of-sight were visually selected from inspection of the {\it IRAS}
100\,$\mu$m emission image of the Pipe nebula prior to the publication by
\citet{LAL06} of the dust extinction map of this cloud complex. In our selection
we chose directions toward high dust emission as well as some directions 
pointing to positions presenting lower emission but close to the main body of the
complex as defined by the 100\,$\mu$m image. After that, \citet{ALL07}
published their list of dense cores and some of our selected fields turned out 
either to completely include one of these cores or part of its outskirts. 
In Fig.\,\ref{pipe_areas} the observed lines-of-sight are overplotted on the 
dust extinction map of the Pipe nebula obtained by \citet{LAL06}. The small 
squares roughly indicates the areas covered by the observed frames.

When in linear polarization mode, the polarimeter incorporates a rotatable, 
achromatic half-wave retarder followed by a calcite Savart plate. The 
half-wave retarder can be rotated in steps of 22\fdg 5, and one polarization 
modulation cycle is covered for every 90\degr\ rotation of this waveplate. 
This arrangement provides two images of each object on the CCD with
perpendicular polarizations (the ordinary, $f_o$, and the extraordinary,
$f_e$, beams). Rotating the half-wave plate by 45\degr\ yields in a rotation
of the polarization direction of 90\degr . Thus, at the CCD area where $f_o$
was first detected, now $f_e$ is imaged and vice versa. Combining all four
intensities reduces flatfield irregularities. In addition, the simultaneous
imaging of the two beams allows observing under non-photometric conditions
and, at the same time, the sky polarization is practically canceled. Eight CCD
images were taken for each field with the polarizer rotated through 2
modulation cycles of 0\degr, 22\fdg 5, 45\degr, and 67\fdg 5 in rotation
angle. 

Among the 46 sky positions, twelve were observed at the IAG 60\,cm 
telescope. At this telescope the integration time was set to 120 seconds 
and 5 frames were collected and co-added for each position of the half-wave 
plate (totalizing 600 seconds per wave plate position). The remaining 34 
fields were observed at the 1.6\,m telescope, where the integration time for 
most of the observed positions was also set to 120 seconds, being that only 
one frame was acquired for each position of the half-wave plate. In order to 
have almost the same field of view, the latter telescope was provided with a 
focal reducer.

The CCD images were corrected for read-out bias, zero level bias
and relative detector pixel response. After these normal steps of
CCD reductions, we identified the corresponding pairs of stars and
performed photometry on them in each of the eight frames of a given field
using the IRAF DAOPHOT package. From the obtained file containing magnitude
data, we calculate the polarization by use of a set of specially
developed IRAF tasks \citep[PCCDPACK package; ][]{AP00}. This set includes
a special purpose FORTRAN routine that reads the data files and calculates 
the normalized linear polarization from a least-square solution, which 
yields the degree of linear polarization (P), the polarization position 
angle ($\theta$ , measured from north to east) and the Stokes parameters 
Q and U, as well as the theoretical (i.e. the photon noise) and measured 
errors. The latter are obtained from the residuals of the 
observations at each waveplate position angle $(\psi_i)$ with
respect to the expected $\cos 4 \psi_i$ curve.

Zero polarization standard stars were observed every run to check for any
possible instrumental polarization, which proved to be small
as can be verified by inspection of Table\,\ref{table:1}. The reference 
direction of the polarizer was determined by observing polarized standard 
stars \citep{TBW90}, complemented with polarized stars from the catalogue 
compiled by \citet{heiles}. For all observing seasons, the instrumental 
position angles showed a perfect correlation with the standard values
(see Table\,\ref{table:2}), and the expected uncertainty of
the zero point for the reference direction must be smaller than 1--2\degr.

\subsection{Results}\label{results}

Our final sample contains 11\,948 stars, being that 9\,777 of them have
${\rm P}/\sigma_P \ge 5$, where $\sigma_P$ means the largest between the
theoretical and measured errors, that is, about 3\,200 
stars more than the ones used in the analysis conducted in \citetalias{AFG08}
which limited the sample to stars presenting ${\rm P}/\sigma_P \ge 10$. A 
search in the archival Two-Micron All-Sky  Survey (2MASS), available on-line 
{\tt http://irsa.ipac.caltec.edu}, identified 
11\,588 objects that could be associated to our observed 
stars, and the $JHK_s$ photometry was retrieved for them. 
For the remaining we usually failed to associate a 2MASS object either because none was found inside the searched box (up to about 8 times the typical rms error
of our astrometric solution) or, in case of the most crowded fields, because the
same 2MASS object could be assigned to more then one of ours.

\begin{figure}
\plotone{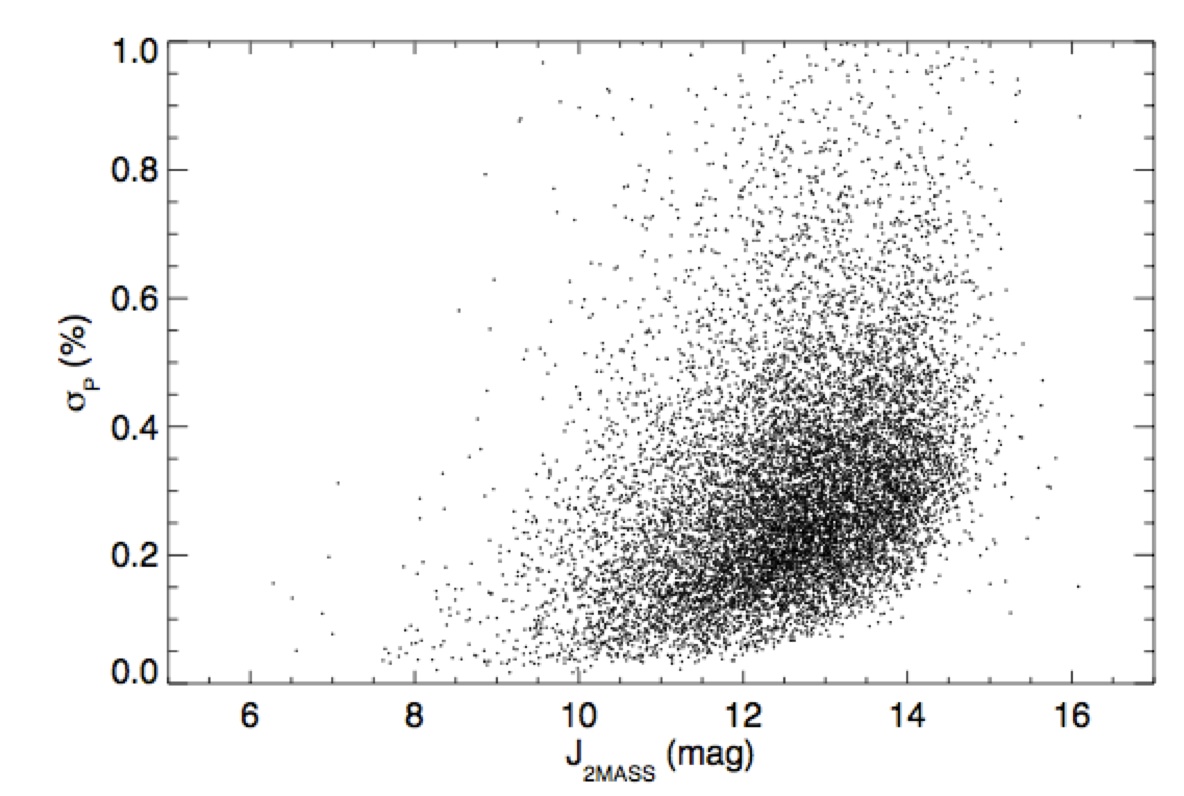}
\caption{Distribution of the polarimetric errors as a function of the $J$ magnitude
as retrieved from the 2MASS catalogue. The distribution shows characteristics of estimated errors dominated by photon shot noise.}
\label{sigma_j}
\end{figure}

Figure\,\ref{sigma_j} gives the distribution of the estimated polarimetric 
errors, $\sigma_P$, as a function of the $J_{\rm 2MASS}$ magnitude, for the 
observed stars. This figure shows that most of our stellar sample has 
magnitude within the interval $10^{\rm m} \le J_{\rm 2MASS} \le 15^{\rm m}$ and 
polarimetric error given by $\sigma_P \le 0.5$\,\%. The obtained distribution
suggests that the uncertainties are dominated by photon shot noise, as 
expected for a sample collected with fixed exposure time.

\begin{figure}
\plotone{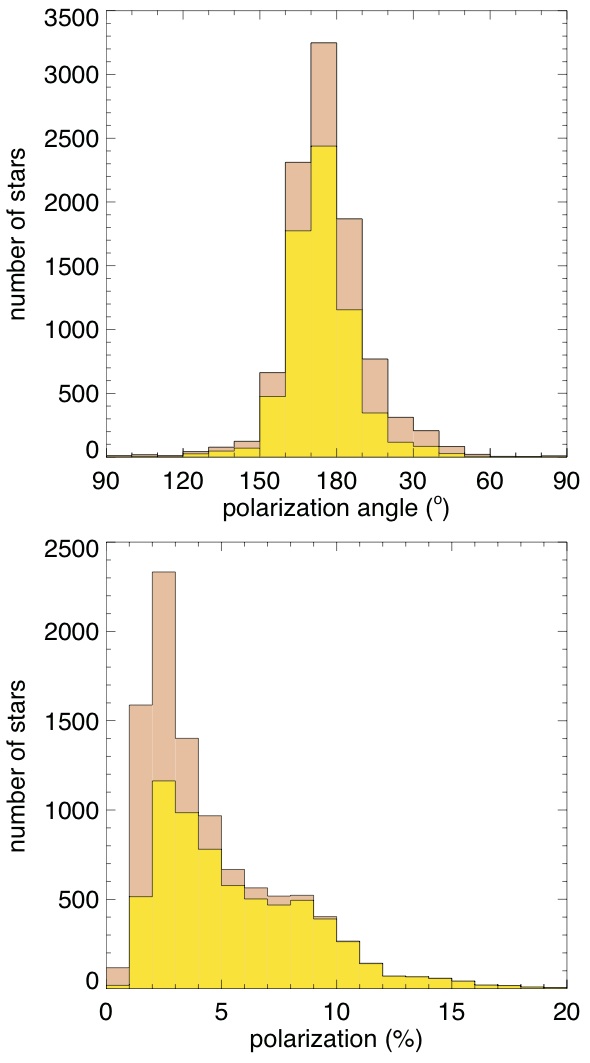}
\caption{The obtained distributions for the 9\,777 stars with $P/\sigma_P \ge 5$
and for the 6\,582 stars with $P/\sigma_P \ge 10$, darker (salmon) and lighter
(yellow) histograms respectively. {\it Top panel:} distribution of the observed 
polarization angles; {\it Botton panel:} distribution of the estimated degree of 
polarization.}
\label{histograms}
\end{figure}

The distribution of obtained degree of polarization for stars with
${\rm P}/\sigma_P \ge 5$ (Fig.\,\ref{histograms} -- {\it bottom panel}) shows
a surprising result: several stars present degree of polarization 
larger than 15\%. Indeed, 6 objects present degree of polarization slightly larger
than 19\%. As far as we know, these are the largest polarization 
produced by dichroic extinction ever observed --- in the stellar polarization
catalogue compiled by \citet{heiles} one find only four stars, out of more than 
9\,000, with degree of polarization higher than 10\%, being that the highest of 
them equals to 12.47\%. 

The distribution of obtained polarization angles for stars with 
${\rm P}/\sigma_P \ge 5$ (Fig.\,\ref{histograms} -- {\it top panel}) shows a 
large concentration of values around $\theta \approx 180\degr$ (in equatorial
coordinates), that is, 76\% of this subsample has polarization angles 
between 160\degr\ and 10\degr, clearly indicating a high large scale 
homogeneity in dust grain alignment (and supposedly in the geometry of the 
magnetic field) all over the whole region. Since the main axis of the Pipe's
{\it stem} is almost in line with the west-east direction, it clearly indicates that 
the observed polarization vectors are mainly perpendicularly aligned to the 
longer axis of the cloud \citepalias[see][]{AFG08}.

It is interesting to compare the distributions for stars having
${\rm P}/\sigma_P \ge 5$ with the ones having ${\rm P}/\sigma_P \ge 10$, also
shown in Fig.\,\ref{histograms}, and used in our previous work. Visually one can
attest that the distribution of polarization angles for both samples are quite
similar. Indeed,  the mean value of both distributions differ by only 2\degr\ and
the standard deviation of the sample having ${\rm P}/\sigma_P \ge 5$ is about
2\fdg5 wider than the one for ${\rm P}/\sigma_P \ge 10$. On the other hand, as
expected for a sample with uncertainties dominated by photon shot noise, most 
of the stars included by the less tight signal-to-noise condition are the ones 
presenting smaller degree of polarization.

The high values of polarization obtained in our survey is the result of differential 
extinction produced by interstellar dust grains on the incoming/background 
stellar radiation. It is assumed that a large fraction of those 
grains are aligned. The nature of such alignment is still a matter of 
debate \citep[see][for a comprehensive review on this subject]{L03}, however, 
it is widely believed that the dominant process responsible for the alignment 
involves interaction between the spin of the dust particles and the ambient 
magnetic field, as originally proposed by \citet{DG51}.

\section{Data Analysis}\label{analysis}

\subsection{Mean Polarization}\label{mean_pol}

Apart from the difference in the adopted signal-to-noise, the results
described in \citetalias{AFG08} and the ones presented in this work were 
obtained in the way described below and introduced in Table\,\ref{table:3} which
gives, for each observed field, coordinates (right ascension and declination), 
associated dark core, when available, from the list compiled by \citet{ALL07}, 
the number of stars for which we estimated polarization, and the number of 
stars with ${\rm P}/\sigma_P \ge 5$, in columns 2 and 3, 4, 5, and 6 respectively.

\begin{deluxetable*}{c c c c c c r r c r c r r c c}
\tablewidth{0pt}
\tabletypesize{\scriptsize}
\tablecaption{Mean $R$-band linear polarization and extinction data for the 46 observed fields in the Pipe nebula (see text for explanation on the columns).\label{table:3}}
\tablehead{Field & \multicolumn{2}{c}{alpha\, (J2000)\, delta} & 
Core\tablenotemark{a} & observed & stars with &
\multicolumn{1}{c}{$\theta_{mean}$\tablenotemark{b}} & \multicolumn{1}{c}{$
\sigma_{std}$\tablenotemark{b}} & stars\tablenotemark{c} & \multicolumn{1}{c}{$
\langle P \rangle$} & \multicolumn{1}{c}{$\delta P$} & \multicolumn{1}{c}
{$\theta_{\langle P \rangle}$} &\multicolumn{1}{c}{$\Delta \theta$} & $A_V$ & 
$\delta A_V$ \\
     & h\,\,\,m\,\,\,s & $\,\,\,\,\circ$\,\,\,\,\,$\prime$\,\,\,\,${\prime\prime}$ & & stars & $P/\sigma \ge 5$ &   \multicolumn{1}{c}{$\circ$} & \multicolumn{1}{c}{$\circ$} &  &  \multicolumn{1}{c}{\%} & \% &  \multicolumn{1}{c}{$\circ$} &  \multicolumn{1}{c}{$\circ$} & mag & mag} 
\startdata
01 & 17\,10\,28 & $-$27\,22\,49 & 06 & 273 & 174 &   9.7 & 10.65 & 168 &  2.45 & 0.82 &   3.1 &  9.65 & 2.54 & 1.25 \\
02 & 17\,11\,52 & $-$27\,03\,49 & --- & 165 &  64 &  12.9 & 15.04 &  63 &  1.82 & 0.85 &   9.8 & 13.72 & 0.63 & 0.35 \\
03 & 17\,11\,21 & $-$27\,24\,46 & 12 & 62 &  23 &  28.9 & 24.93 &  21 &  1.78 & 0.96 &  48.8 & 23.75 & 3.88 & 2.00 \\
04 & 17\,10\,55 & $-$27\,44\,26 & --- & 400 & 211 &  15.9 & 13.08 & 202 &  1.63 & 0.79 &  14.8 & 12.38 & 1.59 & 0.74 \\
05 & 17\,12\,30 & $-$27\,20\,42 & 14& 137 &  50 &   0.9 & 16.82 &  47 &  2.00 & 1.07 & 179.7 & 15.67 & 2.53 & 1.09 \\
06 & 17\,12\,01 & $-$27\,37\,06 & 08 & 271 &  91 & 128.7 & 40.79 &  87 &  0.62 & 1.36 & 149.8 & 40.21 & 2.33 & 1.19 \\
07 & 17\,13\,53 & $-$27\,12\,33 & --- & 206 & 114 & 167.3 & 12.75 & 108 &  1.27 & 0.61 & 169.3 & 11.74 & 2.06 &  0.97 \\
08 & 17\,13\,34 & $-$27\,45\,46 & --- & 807 & 486 &   7.6 &  8.80 & 464 &  1.76 & 0.63 &   9.2 &  7.40 & 1.58 &  0.78 \\
09 & 17\,14\,52 & $-$27\,20\,55 & 21 & 189 & 132 &   6.3 & 10.76 &  54 &  3.44 & 0.98 &   3.5 & 10.04 & 3.03 & 1.49 \\
10 & 17\,15\,25 & $-$27\,18\,09 & --- & 191 & 174 &   4.2 &  7.72 & 164 &  3.43 & 1.23 &   4.4 &  7.29 & 3.33 & 1.67 \\
11 & 17\,15\,15 & $-$27\,33\,38 & 20 & 135 & 130 & 170.8 &  6.76 & 127 &  3.17 & 1.14 & 171.3 &  6.21 & 2.32 & 1.17 \\
12 & 17\,16\,20 & $-$27\,09\,32 & 25 & 198 & 189 & 176.5 &  4.87 & 179 &  4.46 & 1.01 & 176.8 &  4.32 & 2.72 & 1.22 \\
13 & 17\,17\,12 & $-$27\,03\,06 & 27 & 199 & 196 &   0.9 &  5.76 & 190 &  4.03 & 0.69 &   1.3 &  5.62 & 2.03 & 0.94 \\
14 & 17\,16\,05 & $-$27\,31\,38 & 23 & 282 & 280 & 176.0 &  4.40 & 267 &  4.34 & 0.93 & 176.1 &  4.11 & 2.82 & 1.42 \\
15 & 17\,18\,27 & $-$26\,47\,50 & 31 & 272 & 254 &   7.7 & 10.90 & 243 &  2.18 & 0.78 &   5.3 & 10.63 & 2.32 & 1.16 \\
16 & 17\,18\,48 & $-$27\,11\,36 & --- & 382 & 319 &   2.2 &  4.92 & 301 &  2.39 & 0.50 &   2.6 &  3.52 & 1.98 & 0.98 \\
17 & 17\,19\,36 & $-$26\,55\,23 & 33& 214 & 210 & 171.6 &  5.91 & 201 &  2.58 & 0.76 & 171.8 &  5.58 & 2.07 & 0.91 \\
18 & 17\,20\,49 & $-$26\,53\,08 & 34/40 & 327 & 201 & 164.2 &  9.59 & 198 &  4.63 & 1.68 & 163.0 &  8.60 & 3.64 & 1.81 \\
19 & 17\,22\,43 & $-$26\,39\,25 & --- & 368 & 323 & 167.7 &  7.06 & 306 &  2.31 & 0.79 & 167.2 &  6.35 & 1.86 & 0.94 \\
20 & 17\,24\,03 & $-$26\,20\,35 & --- & 451 & 400 &  32.2 &  9.86 & 383 &  2.05 & 0.73 &  32.5 &  9.23 & 1.49 & 0.68 \\
21 & 17\,21\,48 & $-$27\,18\,15 & --- & 260 & 218 & 179.1 &  8.73 & 209 &  1.96 & 0.38 &   0.8 &  8.05 & 2.33 & 1.16 \\
22 & 17\,22\,38 & $-$27\,04\,14 & 41/42 & 102 &  97 & 158.1 &  5.35 &  92 &  4.39 & 1.33 & 157.0 &  4.82 & 3.03 & 1.50 \\
23 & 17\,27\,13 & $-$25\,07\,27 & --- & 513 & 412 &   7.1 &  9.44 & 391 &  2.37 & 0.74 &   7.1 &  8.79 & 2.03 & 0.99 \\
24 & 17\,26\,25 & $-$25\,58\,09 & --- & 748 & 520 & 175.1 &  7.61 & 506 &  2.59 & 0.76 & 174.2 &  6.52 & 2.12 & 1.05 \\
25 & 17\,28\,07 & $-$25\,29\,52 & --- & 511 & 461 & 174.2 &  5.58 & 437 &  3.27 & 0.74 & 173.9 &  4.91 & 2.17 & 1.08 \\
26 & 17\,25\,40 & $-$26\,43\,09 & 48 & 247 & 126 & 142.3 & 32.79 & 120 &  1.99 & 1.49 & 152.9 & 32.52 & 2.21 & 1.26 \\
27 & 17\,25\,28 & $-$27\,03\,29 & --- & 197 & 185 & 143.2 & 11.43 & 183 &  3.39 & 1.30 & 142.2 & 11.19 & 2.21 & 1.10 \\
28 & 17\,30\,18 & $-$25\,09\,50 & --- & 254 & 240 & 172.9 &  5.86 & 231 &  5.61 & 1.16 & 173.3 &  5.35 & 3.18 & 1.59 \\
29 & 17\,29\,14 & $-$25\,55\,44 &  $\sim$70 & 98 &  95 & 160.2 &  5.90 &  91 &  5.43 & 1.52 & 160.8 &  5.46 & 3.59 & 1.76 \\
30 & 17\,28\,12 & $-$26\,21\,10 &  56 & 94 &  93 & 160.6 &  2.54 &  90 &  6.64 & 2.17 & 160.6 &  2.42 & 2.79 & 1.39 \\
31 & 17\,27\,12 & $-$26\,42\,59 & 51 & 284 & 271 & 155.1 &  5.12 & 258 &  4.99 & 1.59 & 156.1 &  4.61 & 3.47 & 1.73 \\
32 & 17\,27\,24 & $-$26\,56\,50 & 47 & 143 & 139 & 164.1 &  4.58 & 135 &  6.20 & 1.54 & 164.5 &  3.92 & 3.81 & 1.88 \\
33 & 17\,32\,09 & $-$25\,24\,18 & 91 & 329 & 313 & 169.9 &  4.04 & 301 &  8.10 & 1.30 & 169.3 &  3.22 & 4.38 & 2.16 \\
34 & 17\,32\,54 & $-$25\,12\,25 & --- & 255 & 244 & 168.9 &  3.17 & 236 &  7.98 & 1.38 & 169.9 &  2.54 & 3.91 & 1.87 \\
35 & 17\,33\,01 & $-$25\,46\,00 & $\sim$89 & 133 & 130 & 171.7 &  2.37 & 125 & 10.83 & 1.80 & 171.3 &  1.95 & 4.48 & 2.23 \\
36 & 17\,30\,11 & $-$26\,48\,42 & --- & 144 & 142 & 160.4 &  3.76 & 138 &  9.79 & 1.62 & 160.1 &  3.65 & 3.24 & 1.51 \\
37 & 17\,31\,18 & $-$26\,29\,36 & 66 & 111 & 111 & 169.8 &  3.40 & 105 & 13.92 & 2.29 & 170.6 &  3.11 & 4.53 & 2.15 \\
38 & 17\,32\,27 & $-$26\,15\,49 & 74 & 127 & 127 & 172.6 &  3.36 & 125 & 15.51 & 2.85 & 173.4 &  3.26 & 4.48 & 2.11 \\
39 & 17\,38\,56 & $-$24\,08\,57 & 151 & 249 & 245 &   6.5 &  6.07 & 233 &  3.87 & 1.04 &   7.8 &  5.83 & 2.29 & 1.10 \\
40 & 17\,35\,47 & $-$25\,33\,01 &  109 & 80 &  77 & 165.2 &  4.07 &  73 & 11.04 & 1.84 & 167.0 &  3.88 & 3.55 & 1.66 \\
41 & 17\,36\,27 & $-$25\,23\,27 &  --- & 62 &  62 & 170.8 &  4.30 &  57 & 10.54 & 2.16 & 170.0 &  4.19 & 2.81 & 1.36 \\
42 & 17\,33\,54 & $-$26\,14\,11 & --- & 181 & 177 & 174.6 &  3.05 & 170 &  9.49 & 1.68 & 174.3 &  2.91 & 3.21 & 1.53 \\
43 & 17\,33\,24 & $-$26\,41\,13 & --- & 424 & 422 & 167.3 &  3.03 & 406 &  8.06 & 2.07 & 167.2 &  2.84 & 3.10 & 1.53 \\
44 & 17\,37\,55 & $-$25\,12\,40 & 132 & 119 & 114 &   0.7 &  3.76 & 108 &  8.49 & 1.51 &   0.2 &  3.59 & 3.65 & 1.79 \\
45 & 17\,39\,50 & $-$24\,59\,16 & 140 & 412 & 401 & 177.9 &  5.51 & 389 &  6.29 & 1.57 & 177.5 &  5.24 & 4.04 & 1.95 \\
46 & 17\,37\,56 & $-$26\,15\,32 & --- & 363 & 353 & 169.8 &  4.13 & 335 &  6.74 & 1.59 & 169.6 &  3.82 & 3.55 & 1.65
\enddata
\tablenotetext{a}{Identification from \citet{ALL07}.}
\tablenotetext{b}{Obtained by a Gaussian fitting to the distribution of observed 
polarization angles (measured from the North Celestial Pole to East).}
\tablenotetext{c}{Number of stars passing the selection criteria which were used 
to estimate the mean values, $\langle P \rangle$ and $\theta_{\langle P \rangle}$,
(see text).}
\end{deluxetable*}

Mean polarization and polarization degree were estimated for each of the 
observed area adopting a procedure similar to the one used by \citet{PM07}, 
that is, to improve the precision of the mean values, we selected only 
those objects with observed polarization angle $\theta_{obs}$ 
within the interval ($\theta_{mean} - 2 \sigma_{std} \le \theta_{obs} \le  
\theta_{mean} + 2 \sigma_{std}$) where, $\theta_{mean}$ and $\sigma_{std}$ 
are the mean polarization angle and standard deviation of each field sample 
(columns 7 and 8, respectively, in Table\,\ref{table:3}). The mean Stokes 
parameters, $\langle Q \rangle$ and $\langle U \rangle$, for each field, 
were estimated from the individual values for each star ($q_i, u_i$), 
weighted by the error ($\sigma_i$) according to 

$$\langle Q \rangle = \frac{\sum (q_i/\sigma^2_i)}{\sum \sigma^{-2}_i}$$
$$\langle U \rangle = \frac{\sum (u_i/\sigma^2_i)}{\sum \sigma^{-2}_i}$$
The estimated mean polarization value $\langle P \rangle$ and its associated 
error $\delta P$ are then given by

$$ \langle P \rangle = \sqrt{\langle Q \rangle^2 + \langle U \rangle^2},$$
$$\delta P = \frac{\delta Q\, |\langle Q \rangle| + \delta U\, |\langle U \rangle|}{\langle P \rangle}$$
where $\delta Q$ and $\delta U$ are the estimated standard deviation for the
mean Stokes parameters $\langle Q \rangle$ and $\langle U \rangle$, 
respectively. The mean polarization position angle $\theta_{\langle
P \rangle}$ is given by
$$\theta_{\langle P \rangle} = 0.5\,{\rm tan}^{-1} \left ( \frac{\langle U \rangle}{\langle
Q \rangle} \right )$$

\begin{figure*}
\epsscale{1.}
\plotone{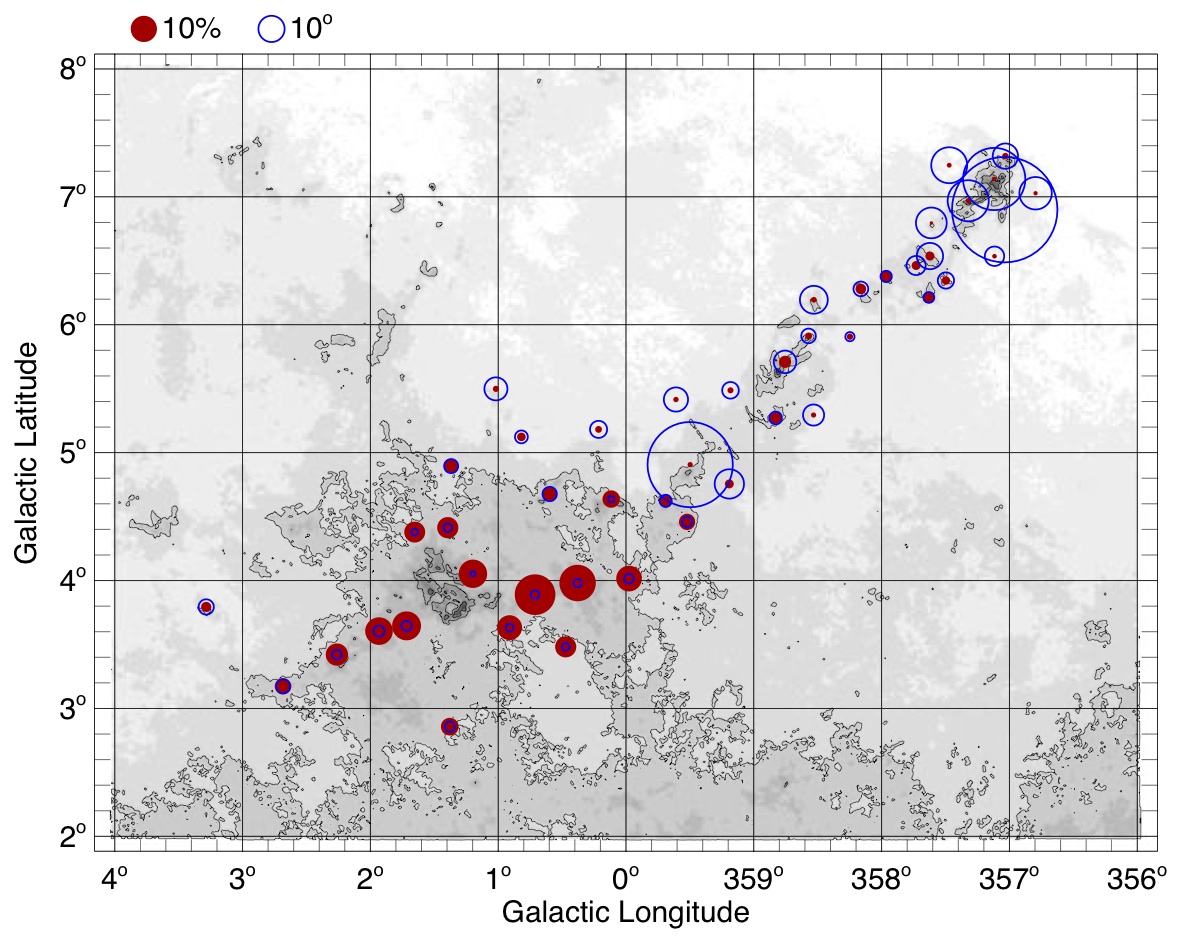}
\caption{Representation of the mean polarization degree (filled circles) and dispersion of polarization angles (open circles) for the observed areas. The size of the symbols are proportional to the scale indicated over the left-hand corner of the image. The anti-correlation between dispersion in polarization angle and mean polarization is clearly seen.}
\label{pol_ang}
\end{figure*}

The number of stars passing the $2\sigma_{std}$ filter, the obtained mean 
polarization, its estimated uncertainty, and the polarization angle for the mean 
polarization vector are given in columns  9, 10, 11, and 12 of 
Table\,\ref{table:3}, respectively. Column 13 shows the dispersion of 
polarization angles corrected in quadrature by the mean error of the
polarization position angle, that is, $\Delta\theta = (\sigma_{std}^2 - 
\langle \sigma_\theta \rangle^2)^{1/2}$, where the mean error, $\langle
\sigma_\theta \rangle$, was estimated from $\langle \sigma_\theta \rangle
= \sum \sigma_{\theta i}/N$, where $\sigma_{\theta i}$ is the 
estimated uncertainty of the star's polarization angle\footnote{The
uncertainty of the polarization angle is estimated by error propagation in the
expression of the position angle $\theta$, which yields $\sigma_{\theta}
= \frac{1}{2} \, \sigma_P/P$, in radians, or $\sigma_{\theta}
= 28\fdg65 \, \sigma_P/P$ \citep[see for instance,][]{SK74} when expressed in
degrees.}.

The global polarimetric properties of the Pipe nebula were already presented in 
\citetalias{AFG08}, and show some interesting results. For instance, the obtained
mean polarizations for the region of B\,59 and along the {\it stem} are typical
for star formation regions \citep[e.g.,][]{VC93, WG94, WG01}, while the values 
obtained for the {\it bowl} are unusually high. Another noteworthy result
presented in \citetalias{AFG08} is the apparent general tendency of decreasing dispersion in polarization angles along the filamentary  structure of the Pipe nebula 
from B\,59 toward the {\it bowl}, while the mean degree of polarization increases 
toward this end. This effect is better visualized by inspection of the image 
presented in Fig.\,\ref{pol_ang}, where we represented the obtained  mean 
polarization degree and dispersion of polarization angles by filled and open circles, respectively, scaled to the values of these observational quantities. In fact, this
figure is more instructive than the diagram introduced in Fig.\,2 of 
\citetalias{AFG08}, because in addiction to the above mentioned anti-correlation
between polarization degree and dispersion of polarization angles seen along
the main axis of the complex, one can also see how these two quantities distributes
spatially over the cloud. In general, one see that fields 
toward lower infrared absorption have the tendency of presenting larger 
values of dispersion of polarization angles. A noticeable exception is Field 26, 
located close to the center of the area displayed on Fig.\,\ref{pol_ang}, 
which presents the second largest  dispersion value in our sample 
($\Delta \theta = 32\fdg5$). 

All fields, but three, having a rather broad distribution of polarization angles 
($\Delta \theta \ge 10\degr$) are in the vicinity of B\,59. The exceptions
are: Field 15, laying in the stem almost middle way from B\,59 to the {\it bowl}, 
the already mentioned Field 26, and Field 27, both located in the eastern side
of the {\it stem}, close to the {\it bowl}. 

\subsection{Polarization maps}\label{pol_maps}

\begin{figure*}
\epsscale{1.}
\plotone{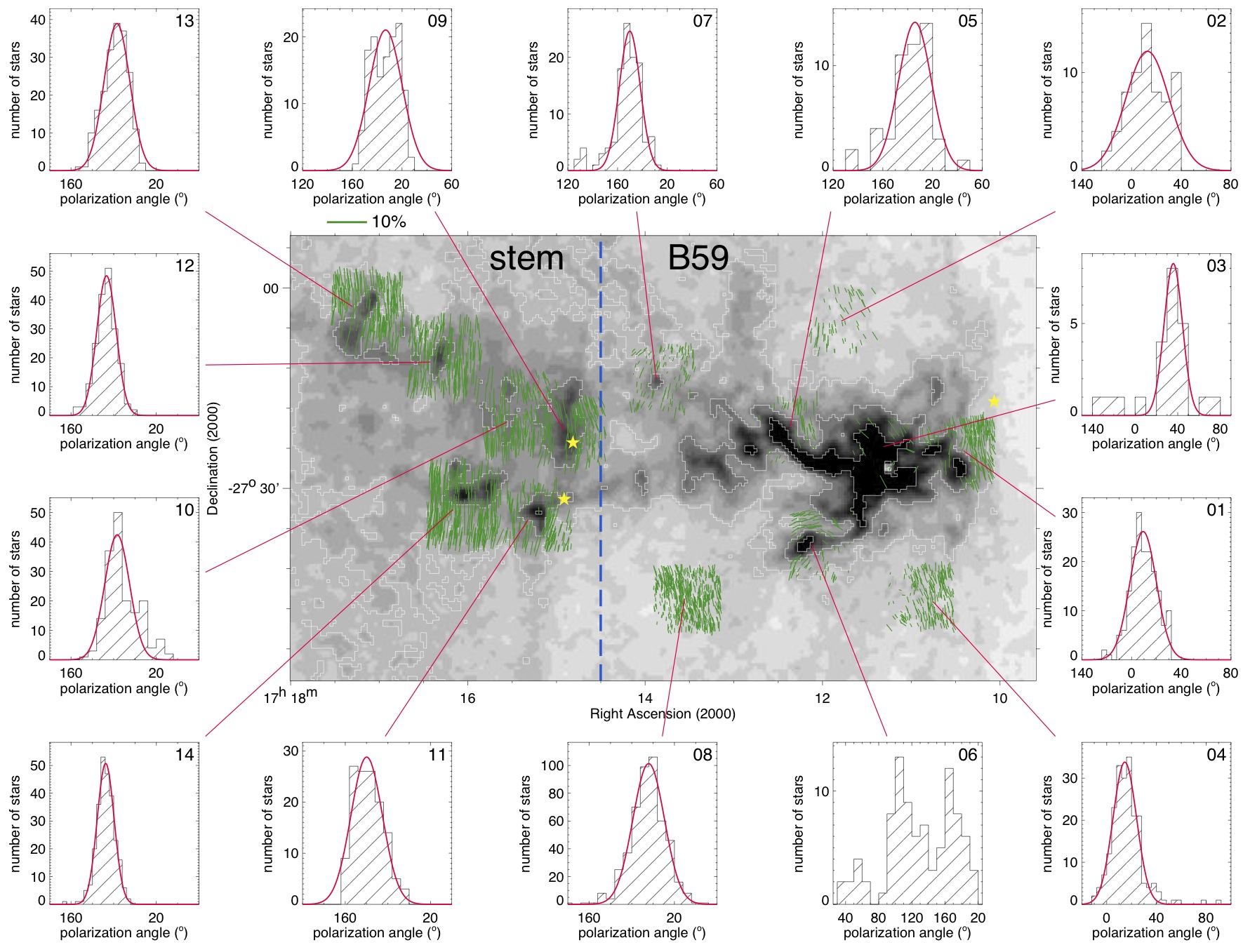}
\caption{Polarization map for Fields 01 to 14 overlapping the dust 
extinction map of the corresponding area \citep{LAL06}. The
overplotted contours are for $A_V =  2, 4$ and $8$ mag. The length 
of the vectors correlates linearly with the degree of polarization 
according to the scale indicated over the left-hand corner
of the image. The vertical dashed-line demarcates the limits between the 
{\it stem} (left) and B\,59 (right), as defined in \citetalias{AFG08}.
Histograms for the distribution of the polarization angles
are shown individually for each field. The identification of the fields
is given in the upper right-hand corner of the histograms. The overplotted
gaussian curves are for comparison purposes only. The `star' symbols indicate the 
location of the identified candidate YSOs by \citet{FLM09}. Note that the 
location of the young stellar cluster identified by \citet{BHB07} in the heart of
B\,59  was omitted. The source on the west side of B\,59 core is the 
Herbig Ae/Be star KK Oph which is very likely associated to the cloud. The two 
other objects at the east of B\,59 are sources 11 (north) and 16 (south) 
listed by \citet{FLM09}.}
\label{field_a}
\end{figure*}

\begin{figure*}
\epsscale{.803}
\plotone{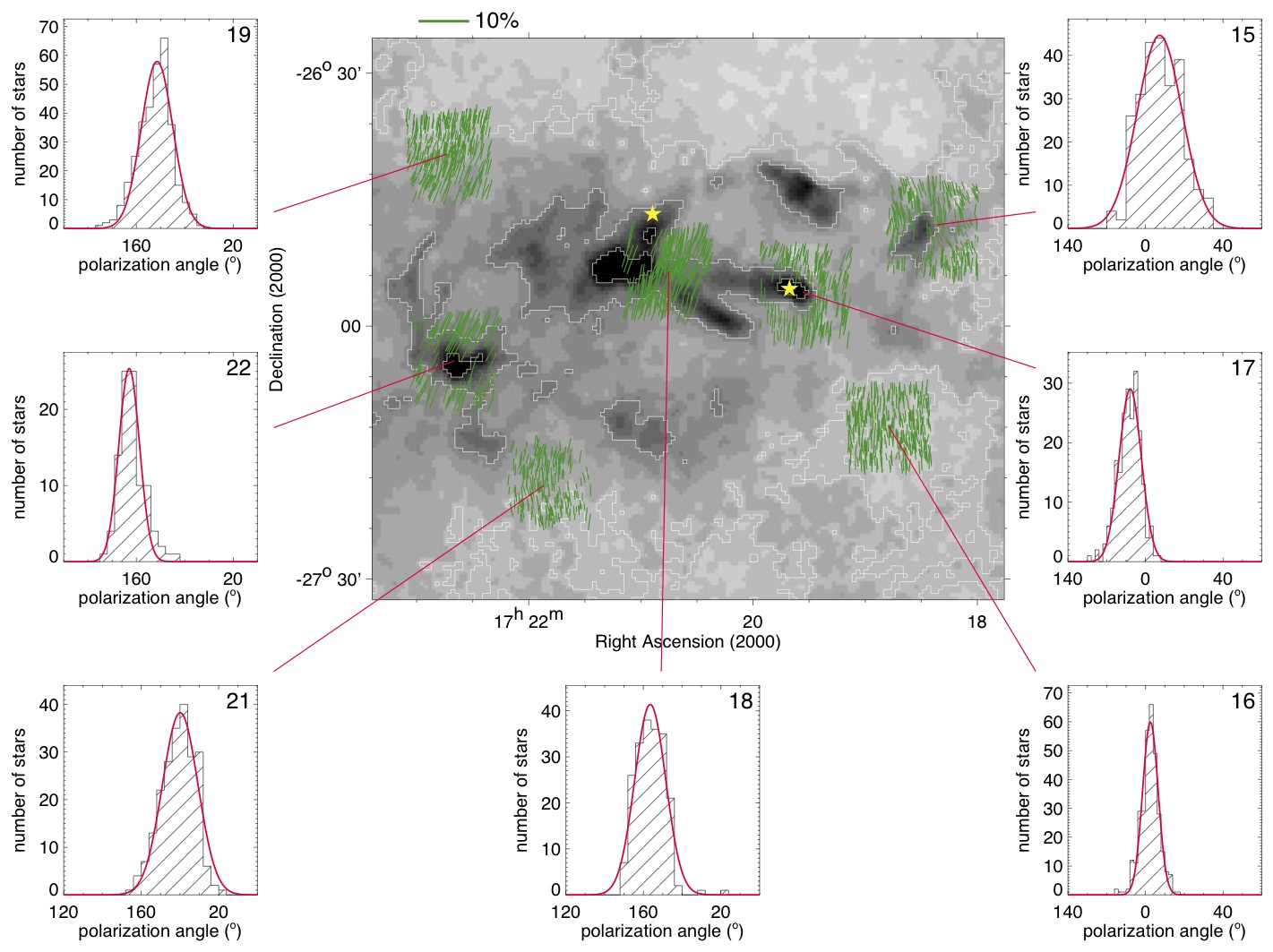}
\caption{Same as Fig.\,\ref{field_a} for Fields 15 to 19, 21, and 22. The `star' 
symbols locate sources 24 (east) and 26 (west) listed by \citet{FLM09}.}
\label{field_b}
\end{figure*}

\begin{figure*}
\epsscale{.937}
\plotone{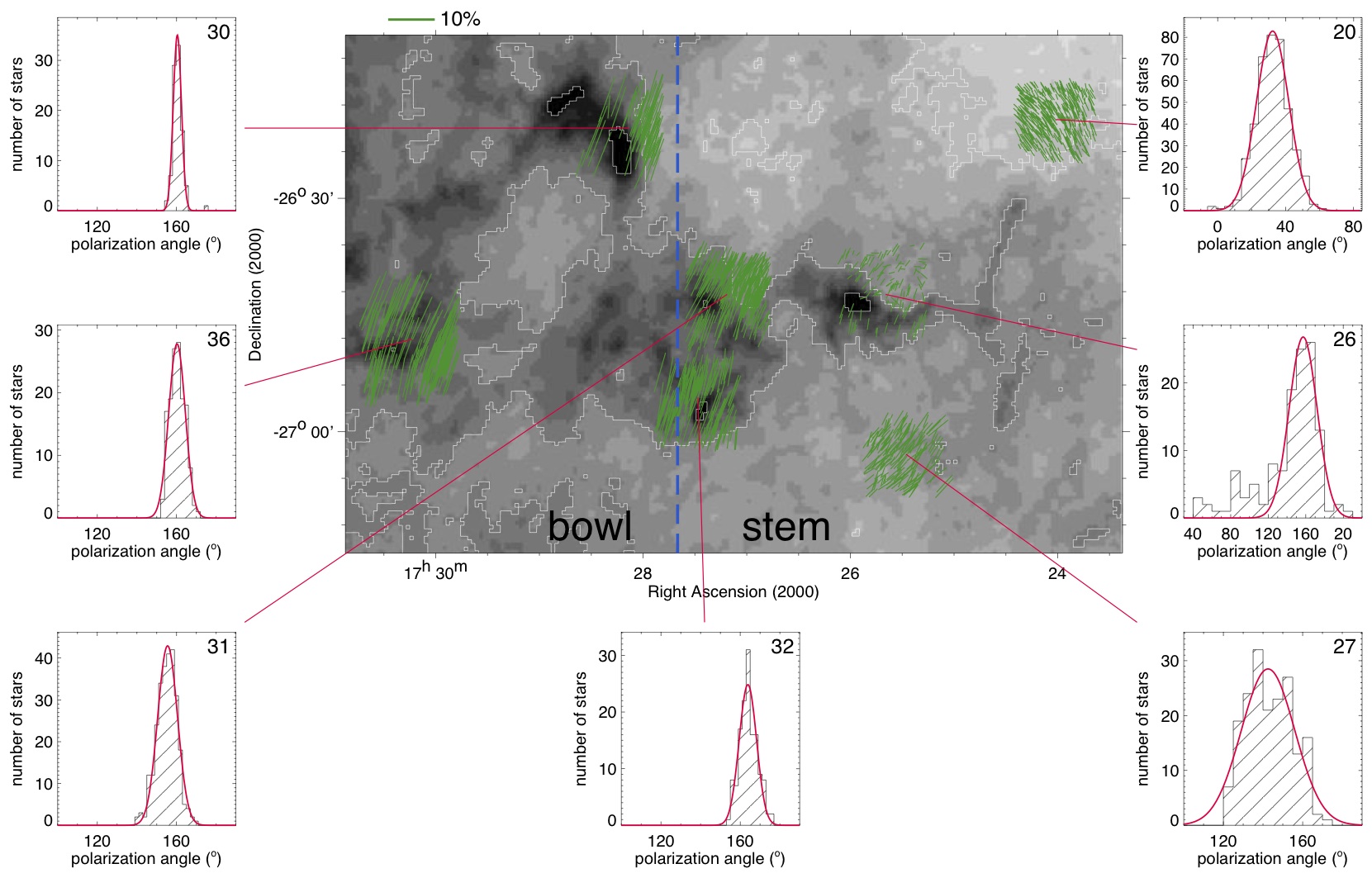}
\caption{Same as Fig.\,\ref{field_a} for Fields 20, 26, 27, 30 to 32, and 36.
The vertical dashed-line demarcates the limits between the {\it bowl} (left)
and the {\it stem} (right), as defined in \citetalias{AFG08}.}
\label{field_c}
\end{figure*}

\begin{figure*}
\epsscale{.6467}
\plotone{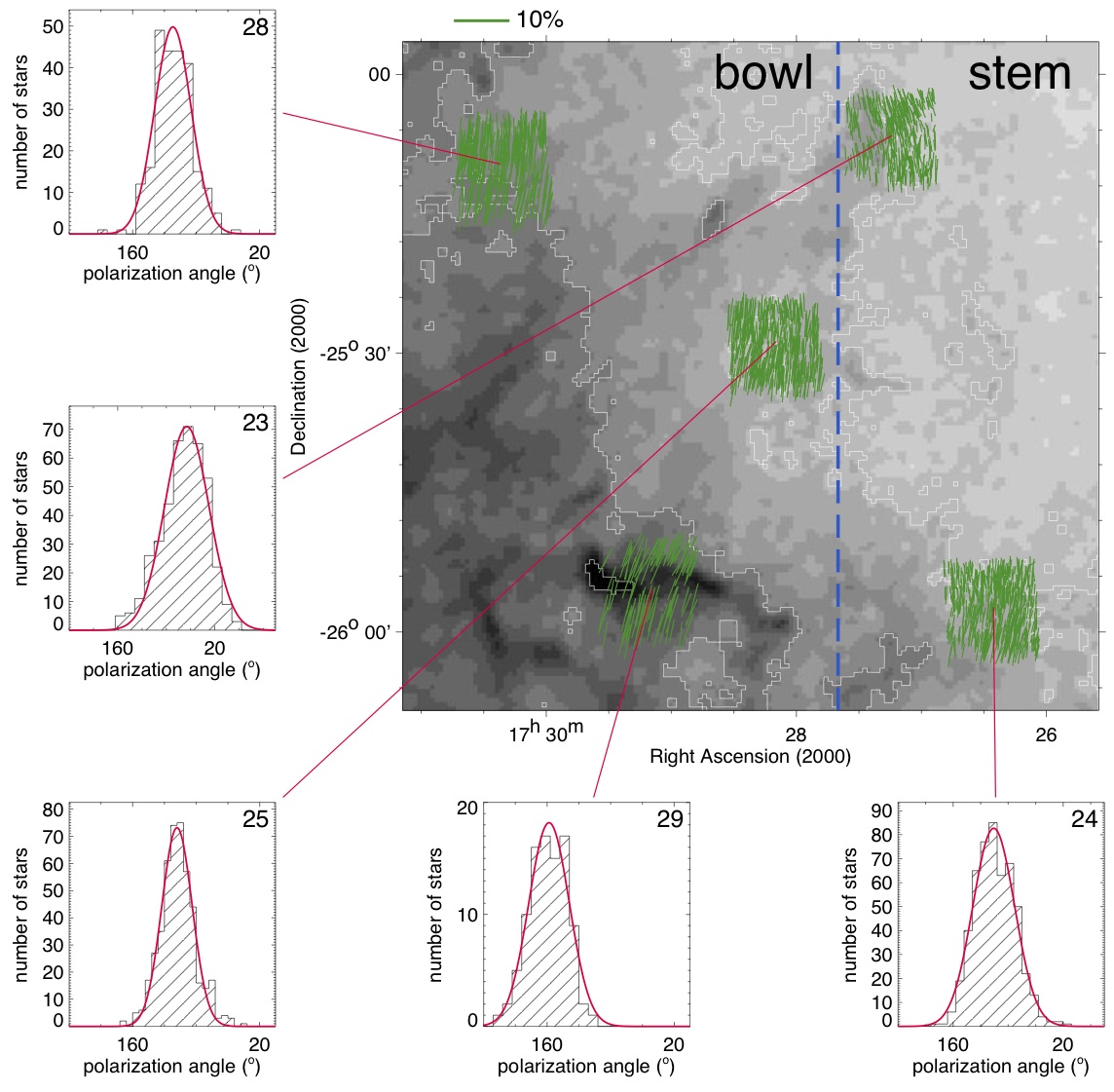}
\caption{Same as Fig.\,\ref{field_a} for Fields 23 to 25, 28, and 29.
The vertical dashed-line demarcates the limits between the {\it bowl} (left)
and the {\it stem} (right), as defined in \citetalias{AFG08}.}
\label{field_e}
\end{figure*}

It is instructive to analyze the obtained polarization for each CCD field.
In Figs.\,\ref{field_a} to \ref{field_d} the obtained polarization is 
overlapped onto the dust extinction maps of the 5 large areas demarcated in 
Fig.\,\ref{pipe_areas}, which cover all observed CCD fields except Fields 39
and 45. The histograms give the distribution of obtained 
position angles for each field, identified in the upper right corner. The 
gaussians represent the best fit to the distribution and are showed for 
comparison purposes only -- they help us to visualize how the distributions
of some fields depart from the ``normal'' distribution. In 
a classical work, \citet{CF53} obtained a reasonably accurate estimate for the 
field strength in the diffuse ISM by directly relating the dispersion in polarization 
position angle to the ratio of two energy densities: the energy density of the 
uniform component of the field and the energy density of turbulence. Since
then, it is widely accepted that the mean value of the distribution of 
polarization position angle obtained from a polarization map gives the angle of 
the mean or uniform (large-scale structured) magnetic field for the region under 
investigation, while the dispersion in the distribution gives information about 
the statistically independent non-uniform (turbulent or random) component of 
the magnetic field \citep[a detailed discussion concerning
this subject can be found in][and references therein]{MG91}. 

The effects of the high interstellar absorption in some of the observed
fields are clearly seen on the distribution of the measured stars. For 
instance, our Field 03, with line-of-sight toward one of the most opaque 
regions of the entire nebula, the B\,59 region \citep{RZ07, RZ09}, is the observed 
field with the smallest number of stars with $P/\sigma_P \ge 5$ (21 stars
only). Its histogram of observed polarization position angles and the  obtained 
mean position angle, $\theta_{\langle P \rangle} = 48\fdg8$ (Table\,\ref{table:3}), 
indicate that most of those stars belong to the right-hand tail of the polarization 
position angle distribution given in Fig.\,\ref{histograms} ({\it top panel}). 
Although the obtained large dispersion of position angles -- which is due to 6 
stars -- the distribution of the remaining stars is rather narrow, as seen in the 
histogram for Field 03 shown in Fig.\,\ref{field_a}. The two polarization angles 
on the right-hand side of the main distribution ($\theta = 64\fdg9$ and $70\fdg1$) 
correspond to [BHB2007] 2 and [BHB2007] 1, respectively, supposed to be 
candidate young stars \citep{BHB07, FLM09}. It is noteworthy that high
resolution optical images of the region show a ``light cone-shaped'' which 
apparently emanates from these stars and illuminates the surrounding dust 
material. Interestingly, both observed polarization vectors are almost perpendicular 
to the symmetry axis of this cone.  

From the histograms shown in Fig.\,\ref{field_a} we note that most of the fields
presenting large dispersion of polarization angles suggest a multicomponent 
structure, in special, Field 06 presents a very interesting geometry for the 
obtained distribution of the polarization vectors and deserves further comments 
(see \S\,\ref{ind_field}). Fields 01 to 04 show distributions of polarization 
angles with many stars having values between 0\degr\ and 40\degr, while the
remaining fields given in Fig.\,\ref{field_a}, already show distributions with
polarization angles between 160\degr\ and 20\degr, likely what was obtained 
for most of the other observed fields.  

The area covered by Fig.\,\ref{field_a} contains, apart from the young stellar
cluster identified by \citet{BHB07} embedded in B\,59, three of the six candidate
YSOs found by \citet{FLM09}. One of them is the well known KK Oph, a 
pre-main-sequence binary with 1.6'' separation and suggested to constitute a 
Herbig Ae star with a classical T Tauri companion \citep[e.g.,][and references 
therein]{HGH05, CAH07}. Although \citet{dWT90} attribute a distance of 310\,pc
to this star, it is commonly accepted a distance of 160\,pc \citep[e.g.,][]{HSV92}
suggesting that this object may have been formed by material formerly 
associated to the Pipe nebula. \citet{CAH07} estimate an age of about 7\,Myr to
this system, that is, from 5 to 6 Myr older than the estimated age of the YSOs in
B\,59. The two other objects are sources 11 and 16 in the \citet{FLM09} candidate 
YSOs list, and have lines-of-sight toward Fields 09 and 11, respectively, close to
the transition between the B\,59 and the {\it stem} regions, as defined in 
\citetalias{AFG08}. These sources were spectroscopically studied by 
\citet{CLR10}, who confirmed the youthful character of the latter, and found that it 
is a visual binary, while the former presents an ambiguous spectra, that is, it
may either be a rather young object or a reddened giant/subgiant. 

Figure\,\ref{field_b} displays the middle portion of the {\it stem}. We observed seven
fields in this area. The obtained histograms seem to present a kind of transition
between the characteristics observed for the B\,59 region and the ones for
the {\it bowl}. That is, Field 15 (one of the fields with $\Delta \theta > 10\degr$)
have a distribution that resembles the ones obtained for the fields in B\,59, 
while very close to it one see Field 16 which shows a distribution with a 
dispersion typical of the ones presented by fields in the {\it bowl}, however,
centered around 0\degr . On the left-hand side of this figure there is Field 22
showing polarization properties with all the characteristics observed in the {\it bowl},
that is, low dispersion of polarization angles and centered around 160\degr.
Two of the \citet{FLM09} YSO candidates are located in this area. One of them
very close to the center of our Field 17 (source 26), the other one close to the
border of Field 18 (source 24). None of these sources were studied by 
\citet{CLR10}, who on the other hand, investigated two other sources that 
turned out to be OH/IR stars, likely residing in the Galactic Bulge.

Figure\,\ref{field_c} displays the ``transition region'' between the {\it stem} and the 
{\it bowl}, as denoted by the vertical dashed-line. In this area we find two 
of the three fields presenting broad distribution of polarization angles not 
belonging to the B\,59 vicinity, Fields 26 and 27, the former shows a distribution of
polarization vectors that resembles the one observed for Field 06, suggesting that
there may be some similarities between the physical properties of both cores,
while the distribution for the latter clearly shows a bimodal distribution of polarization
vectors. All fields presenting particularly interesting polarization distribution are 
separately discussed in \S\,\ref{ind_field}. The four eastern fields of this area
present polarimetric characteristics of the {\it bowl}, that is low dispersion of 
polarization angles, rather high polarization degree and a mean polarization angle 
centered around 160\degr. A detail that calls our attention is the polarization 
probed by Field 20 (upper right corner of the figure). While all other fields shown 
in this figure present a distribution of polarization angles centered around 
$\sim$160\degr, the distribution of polarization angles for Field 20 is centered
around $\sim$30\degr. This is the second less absorbed field 
($A_{\rm V} \approx 1\fm5$), so that the polarization mapped by these stars may
be mainly caused by a background medium.  

The area displayed in Fig.\,\ref{field_e} is located to the north of the one displayed 
in Fig.\,\ref{field_c} and covers mostly the more diffuse medium of the Pipe 
nebula, except for Field 29 with line-of-sight toward a portion with higher 
extinction. Although this field presents a rather large dispersion of position 
angles, as compared to the other fields in the {\it bowl}, its rather high mean 
polarization and mean polarization angle centered around 160\degr, are 
characteristics of that part of the complex.   

Figure\,\ref{field_d} displays eleven fields observed in the {\it bowl} area. The main 
characteristics of the fields observed toward this region of the Pipe nebula are
the high degree of polarization and the highly aligned polarization vectors, as
testified by the low dispersion of polarization angles shown by the histograms 
displayed on this Figure.

It is also interesting to analyse the distribution of mean polarization angle as
a function of the right ascension of the observed fields. Such distribution is
shown in Fig.\,\ref{meanangle}, and as already mentioned, due to a fortuitous 
coincidence this celestial coordinate correlates quite well with the field's 
position along the long axis of the Pipe nebula. Most of the obtained mean 
polarization angles are in the interval $\theta_{\langle P \rangle} \sim 
180\degr \pm 20\degr$, indicating that the local uniform magnetic field is 
somewhat aligned perpendicularly to the main axis of the cloud complex. 

Apart from four fields, identified in Fig.\,\ref{meanangle}, the distribution of
the remaining mean polarization position angles seems to follow a pattern. 
The values obtained for fields toward directions having 
lower infrared absorption, represented by open dots, present a rather 
constant value all over the {\it stem} of the Pipe, including the B\,59 region, 
except for two of the four mentioned fields (Fields 20 and 27). 
On the other hand, the distribution shown by fields with
rather large infrared absorption, represented by filled dots, is more 
interesting. Again, apart from the other two identified fields (Fields 03 and 06, 
both in the B\,59 region), which as we have mentioned earlier show some kind 
of peculiar characteristic, one see that the mean polarization angles for these 
fields seems to be rather constant ($\theta_{\langle P \rangle} \sim 180\degr$) 
from B\,59 to almost the center of the {\it stem}, then decrease slowly until close 
to our arbitrary border of the {\it bowl} region, and rise up again by a small 
value and became almost constant ($\theta_{\langle P \rangle} \sim 170\degr$)
in the {\it bowl}.  

This behaviour somehow suggests that the uniform component of the magnetic
field is ``uniform'' in the surrounding diffuse medium but presents small 
systematic variations along the dense parts of the complex. A remarkable point
to be noted is the fact that the right ascension of Fields 20 and 27 somehow
coincides with the one where the mean position angle of the fields associated to
cores seems to reach its smallest value. Unfortunately our observational data
do not allow us to investigate further this coincidence.  

\subsection{Deriving A$_V$ from 2MASS data}\label{mean_Av}

\begin{figure*}
\epsscale{.98}
\plotone{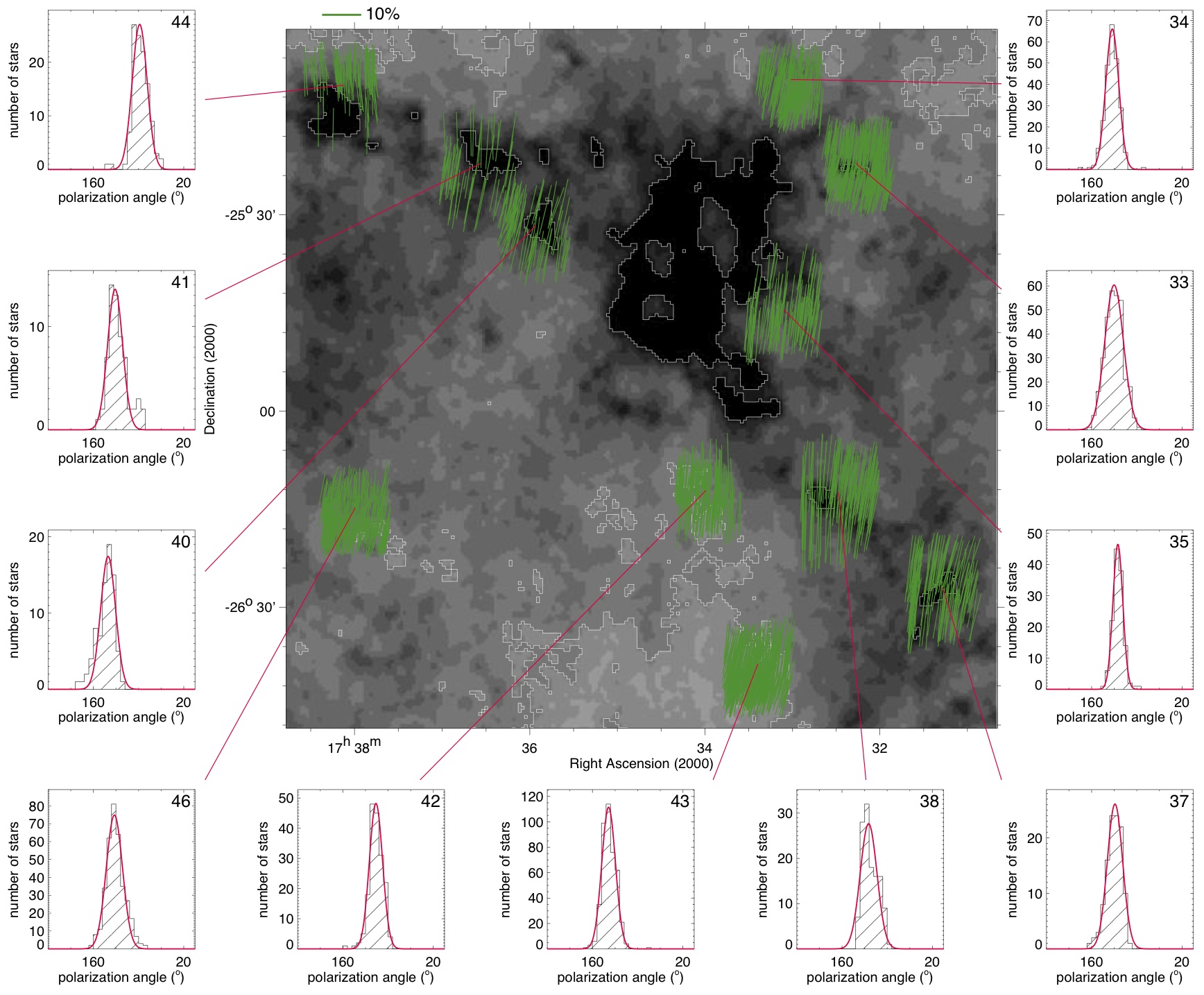}
\caption{Same as Fig.\,\ref{field_a} for Fields 33 to 35, 37, 38, 40 to 44, and 46.}
\label{field_d}
\end{figure*}

It is instructive to compare the obtained mean polarization with
the interstellar extinction acting on each observed line-of-sight. 
The mean extinction in each field could have been estimated
by the use of the extinction map obtained by \citet{LAL06}. In fact, we started
using their image with this purpose, however, as already mentioned, some
of the observed fields contain areas of high interstellar absorption that were 
not probed by our stellar sample. Thus, simply averaging the infrared extinction
over the observed field would provide a larger value for the reddening than the 
one actually probed by our obsereved stars. Because of that, we decided to 
use another approach, that is, the mean extinction in each field may be 
determined by assuming that 
the old bulge population present in each observed volume has an upper 
giant branch similar to that found in $K$, $J-K$ color magnitude diagrams 
(CMD) of Baade's window \citep[see e.g.,][]{FTK99}. We proceeded assuming 
that the upper giant branch in each of our observed fields is comparable to and 
has the same slope as the extinction-corrected template derived by 
\citet{DSB02}, given by 

\begin{equation}
(K_S)_0 = -7.81 \times (J-K_S)_0 + 17.83
\label{gb_locus}
\end{equation}
This assumption is perfectly justified, because those authors applied this
template to study the interstellar reddening in a volume that partially
contains the Pipe nebula. We also assumed that the relation between 
extinction and reddening is given by
\begin{equation}
A_{K_S} = 0.670 \, E(J-K_S).
\label{red_vec}
\end{equation}

\begin{figure}
\epsscale{1.}
\plotone{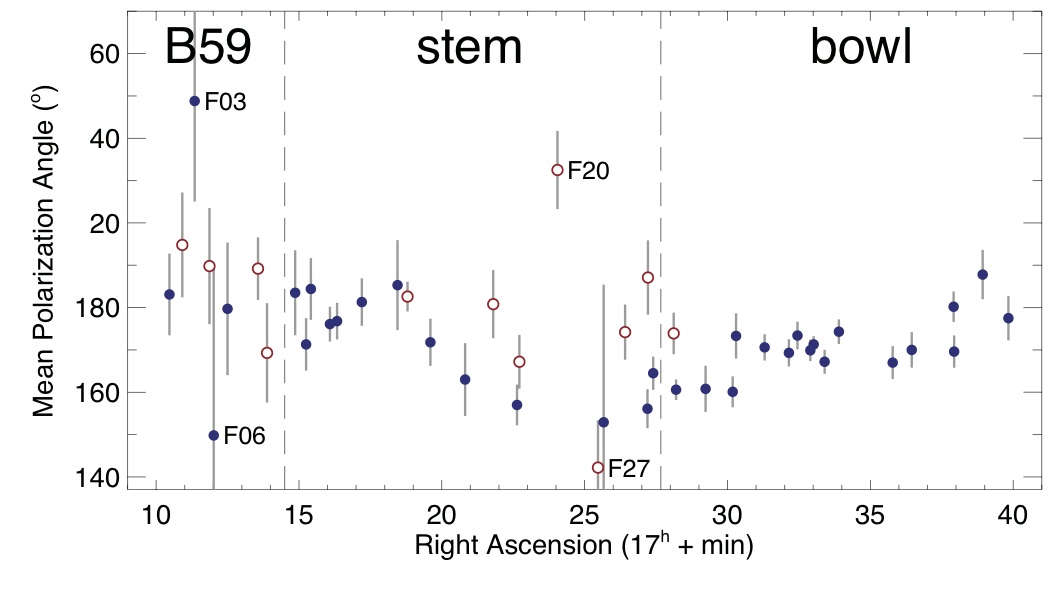}
\caption{Distribution of mean polarization angle, $\theta_{\langle P \rangle}$, as 
a function of the right ascension of the observed field, which correlates quite 
well with its position along the long axis of the Pipe nebula. Filled (blue) and open 
(red) dots represent values for fields associated to lower 
and higher infrared absorptions, respectively. The former are mostly fields 
outside the main structure of the Pipe nebula, namely, Fields 02, 04, 07, 08, 16, 
19 to 21, 23 to 25, and 27. The gray bars give the interval defined by 
$\theta_{\langle P \rangle} \pm \Delta\theta$, where $\Delta\theta$ is the 
dispersion of polarization angles (see Table\,\ref{table:3}).}
\label{meanangle}
\end{figure}

\begin{figure}
\plotone{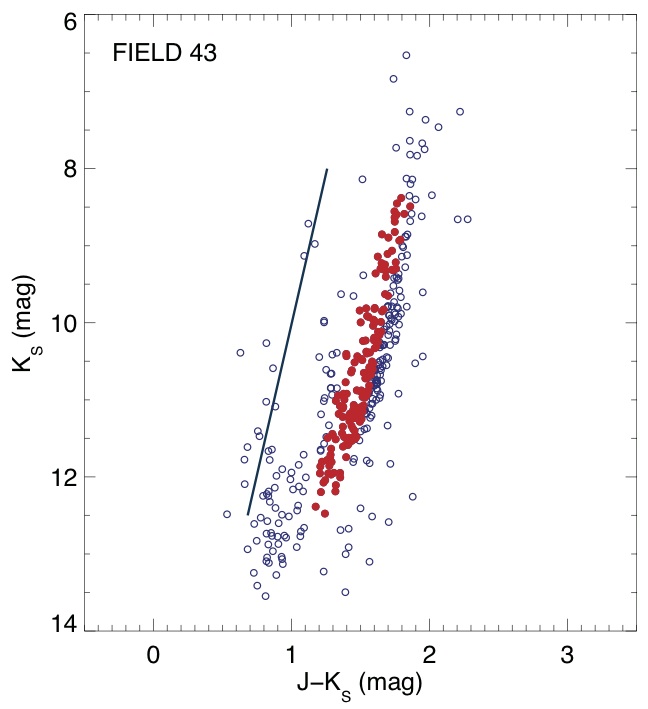}
\caption{Color-magnitude diagram for stars in Field 43. Stars passing the
selection criteria (see text) and used to estimate the field's mean 
interstellar absorption are represented by filled (red) circles. The straight 
line represents the reference upper giant branch (equation \ref{gb_locus}). 
It is clear the gap between the standard reference line and observed stars 
brighter than $K_S \approx 12^{\rm m}$, which corresponds to an
interstellar reddening of about $E(J-K_S) = 0\fm5$.}
\label{Ak_field43}
\end{figure}

From the $(K_S, J-K_S)$ CMD values of each star in the field, we calculated 
the shift along the reddening vector given by Eq.\,(\ref{red_vec}) required 
to make it fall onto the reference upper giant branch, Eq.\,(\ref{gb_locus}). 
Since the adopted template appropriately describes the upper giant branch 
locus for stars with $8 \le (K_S)_0 \le 12.5$, all star presenting 
a corrected $K_S$ magnitude outside this range were excluded from our mean
absorption estimate, and similarly to what we have done when estimating
the mean polarization, a 2-$\sigma$ filter was applied to the obtained distribution
of $E(J-K_S)$ and the field's extinction value was taken as the median of the 
distribution of stars passing the clipping selection. The estimated mean $A_{K_S}$  
values were then converted to $A_V$ by the relation derived by \citet{DSB02}, i.e. 
$A_{K_S}/A_V = 0.118$. To illustrate the method used to estimate the mean 
interstellar absorption, we show, in Fig.\,\ref{Ak_field43}, the CMD
obtained for our Field 43. It is clearly noticeable that most stars brighter 
than $K_S \approx 12^{\rm m}$, in this field, are reddened by about 
$E(J-K_S) = 0\fm5$. Stars used in our estimate of the mean interstellar 
absorption are represented, in Fig.\,\ref{Ak_field43}, by filled circles. In general,
an analysis of the ($J-H$, $H-K_S$) color indices shows that most of 
the stars fainter than $K_S \approx 12^{\rm m}$ in our diagrams are likely to 
be main-sequence stars of earlier types (typically B--G). 

\begin{figure}
\plotone{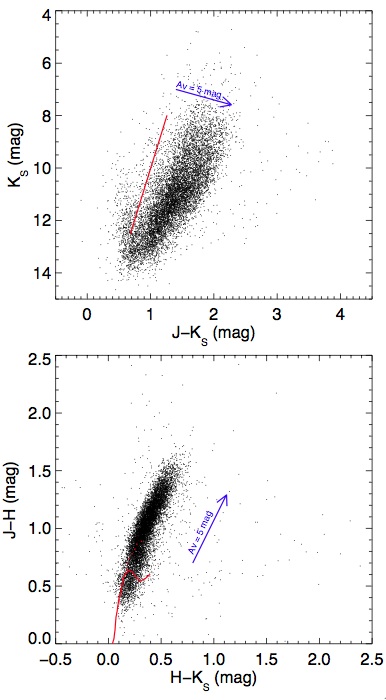}
\caption{{\it Top panel}: $K_S - (J-K_S)$ CMD for all observed star with
$P/\sigma_P \ge 5$ identified in the 2MASS catalogue. The straight line
represents the reference upper giant branch, Eq.\,\ref{gb_locus}, obtained
by \citet{DSB02}. {\it Botton panel:} $(J-H) - (H-K_S)$ color-color 
diagram for the same stars. The theoretical locus for the main-sequence stars
is represented by the continuous line, while the white (red dashed-line) 
represents the giant branch stars. The absorption vector indicated in both 
diagrams corresponds to $A_V = 5^{\rm m}$.}
\label{jhk}
\end{figure}

The top-panel of Fig.\,\ref{jhk} shows the obtained CMD for all observed star,
in our sample, with $P/\sigma_P \ge 5$ and identified in the 2MASS catalogue. 
For comparison, the $(J-H)-(H-K_S)$ diagram for the same stars is given in 
the bottom-panel. As one can see, most of the observed stars seems to 
occupy the area corresponding  to normal reddened stars. The stars in this 
zone could also be dereddened onto intrinsic color lines by extrapolation 
from the observed stellar locus along the appropriate reddening vector.

The obtained values of $A_V$ and their estimated uncertainties 
are given respectively in columns 14 and 15 of Table\,\ref{table:3}. 
The later were estimated from the standard deviation of the obtained 
distribution of $E(J-K_S)$, before applying the 2-$\sigma$ filter, being that the
stellar photometric errors have not been taken into account.

Although the extinction can reach very high values toward some of the 
observed cores, one see in Table\,\ref{table:3} that, as expected, our optical 
polarimetric survey is probing the less absorbed areas only (e.g., from 
$A_V \ge 0\fm6$ to $A_V \le 4\fm6$). It is important to note that although 
\citet{RZ07} found evidences that the extinction law prevailing in the 
densest regions of B\,59 agrees more closely with a dust extinction with a total 
to selective absorption $R_V = 5.5$ we adopted the typical values for the 
diffuse interstellar medium, since we are studying regions with extinctions well 
below $A_K \la 2^{\rm m}$. 

As an independent control of the method used to evaluate 
the interstellar absorption towards the observed areas, one may compare 
the results obtained for our example field, Field 43, with the ones we would 
have obtained from the extinction map produced by \citet{LAL06}. As one may 
verify from visual inspection of Fig.\,\ref{field_d}, our stellar sample covers rather 
uniformly the area defined by Field 43 which means that, simply averaging the 
\citeauthor{LAL06} extinction over this area will probably yield values that are
representative of the mean extinction probed by the observed stars. Such
procedure give us $A_V = 3\fm28 \pm 1\fm26$, which agrees quite well with 
the value given in Table\,\ref{table:3}.  

\section{Polarizing efficiency toward the Pipe nebula}\label{PxAv}

The degree of polarization produced for a given amount of extinction 
is referred to as the ÔÔpolarization efficiencyÕÕ of the intervening dust grains. 
This efficiency of polarization depends on both the nature of the grains and 
the efficiency with which they are aligned in the line-of-sight. The most efficient
polarization medium conceivable is obtained by modeling the dust grains as 
infinite cylinders (very long in comparison to their radii) with diameters 
comparable to the wavelength, perfectly aligned with their long axes parallel 
to one another and perpendicular to the line of sight. For such a model, 
Mie calculations for particles with dielectric optical properties, place a 
theoretical upper limit on the polarization efficiency of the grains due to 
directional extinction at visual wavelengths of  $p/A_V \le 14$\,\%\,mag$^{-1}$  
\citep[see, for instance,][]{whittet2003}. The observations, however, show that 
the upper limit predicted by this very idealized scenario is far from being 
reached. In general, studies of interstellar polarization demonstrate that the 
efficiency of the real Galactic interstellar dust as a polarizing medium is more than 
a factor of 4 less than predicted for the ideal polarizing medium. The observational
upper limit on the ratio of polarization to extinction for the diffuse interstellar
medium is given by $p/A_V \approx 3$\,\%\,mag$^{-1}$ \citep{SMF75}.

\begin{figure}
\plotone{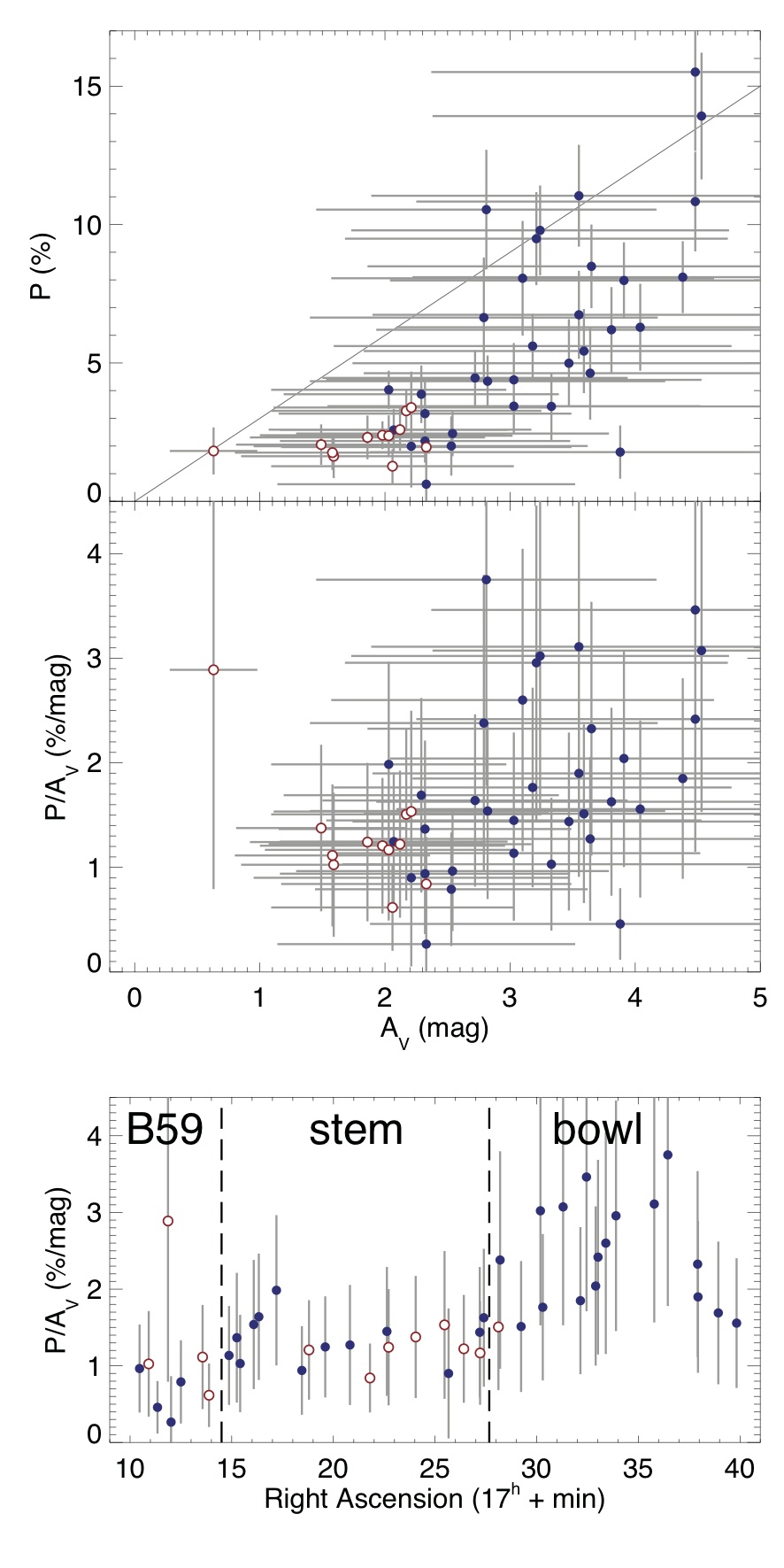}
\caption{{\it Top panel}: Plot of mean polarization ($P$) versus total visual 
absorption ($A_V$) derived  from the 2MASS data for the observed stars with 
$P/\sigma_P \ge 5$. The solid line represents optimum alignment efficiency 
($P(\%) = 3 \times A_V$). {\it Middle panel}: Polarization efficiency ($P/A_V$) 
versus visual absorption $A_V$. {\it Bottom panel:} Distribution of the polarization
efficiency as a function of the right ascension of the observed field. Symbols
have the same meaning as in  Fig.\,\ref{meanangle}.}
\label{polak}
\end{figure}

Considering that our sample contains a rather large number of objects, in the
{\it bowl} region, showing outstanding degrees of polarization, it is natural that we 
try to investigate if these observed areas present unusual polarimetric properties
as compared to the common Galactic interstellar medium.  

The diagrams shown in Fig.\,\ref{polak} were constructed in order to 
investigate the obtained ratio between our estimated mean degree of 
polarization and mean total interstellar absorption for the observed
fields toward the Pipe nebula --- error bars were omitted in these diagrams for the
sake of clarity. The plot of mean polarization {\it versus} total visual absorption 
given in the Fig.\,\ref{polak} ({\it top panel}) shows that basically all data points 
lie on or below the line representing the usual relation 
$p/A_V \approx 3$\,\%\,mag$^{-1}$ --- the two points appearing above 
this line represent data obtained for Fields 38 and 41, however, taking 
into account the estimated 1-$\sigma$ uncertainties for the mean degree 
of polarization and total interstellar absorption, these two fields may also 
obey the above relationship. On the one hand, this result indicates
that the interstellar material composing the Pipe nebula follows the
usual behaviour of the common diffuse interstellar medium. On the other hand, 
as one can see from the values tabulated in Table\,\ref{table:3} we have found 
levels of mean degree of polarization that are unusual for the same interstellar
material. 

Several previous investigations have suggested that the polarizing 
efficiency of the interstellar dust declines  systematically with total 
extinction, as one probes progressively denser environments within a 
dark cloud \citep[e.g.,][]{GJL92, GJL95, GWL95}. The obtained diagram 
of polarizing efficiency, $p/A_{\rm V}$, as a function of the interstellar 
absorption (Fig.\,\ref{polak} -- {\it middle panel}), does not show clearly 
this tendency, at least not for the covered interval of interstellar absorption. 
In fact, on the contrary, if we exclude Field 02, which shows the lowest
interstellar absorption and a polarization efficiency of almost 3\,\%\,mag$^{-1}$, 
the other fields show a tendency of increasing efficiency with the interstellar 
absorption.

More interestingly is the diagram shown in the {\it bottom panel} of Fig.\,\ref{polak},
which shows the distribution of the estimated polarization efficiency of the
observed fields as a function of their position along the long axis of the Pipe
nebula. It is known that variations in polarization efficiency might result from 
changes in physical conditions that affect alignment efficiency, such as 
temperature, density and magnetic field strength, or in grain properties such 
as their shape and size distribution and the presence or absence of surface 
coatings. Most of the observed fields in the {\it stem} (including its tip --- the
B\,59 region), present a polarization efficiency around 
$p/A_V \sim 1$\,\%\,mag$^{-1}$, then it rises up and down when one 
move along the {\it bowl} from west to east, reaching values of about 
$p/A_V \la 4$\,\%\,mag$^{-1}$. Summarizing, although showing an 
interesting behaviour, the global properties of the probed dust material 
composing the Pipe nebula does not seem to present any special peculiarity, 
when compared to the common diffuse interstellar medium, that could explain the
observed high degrees of polarization. However, one notice a clear difference 
between the behaviour shown by the polarimetric properties presented by 
fields located in the {\it stem} and in the {\it bowl}. 

\begin{figure}
\epsscale{1.}
\plotone{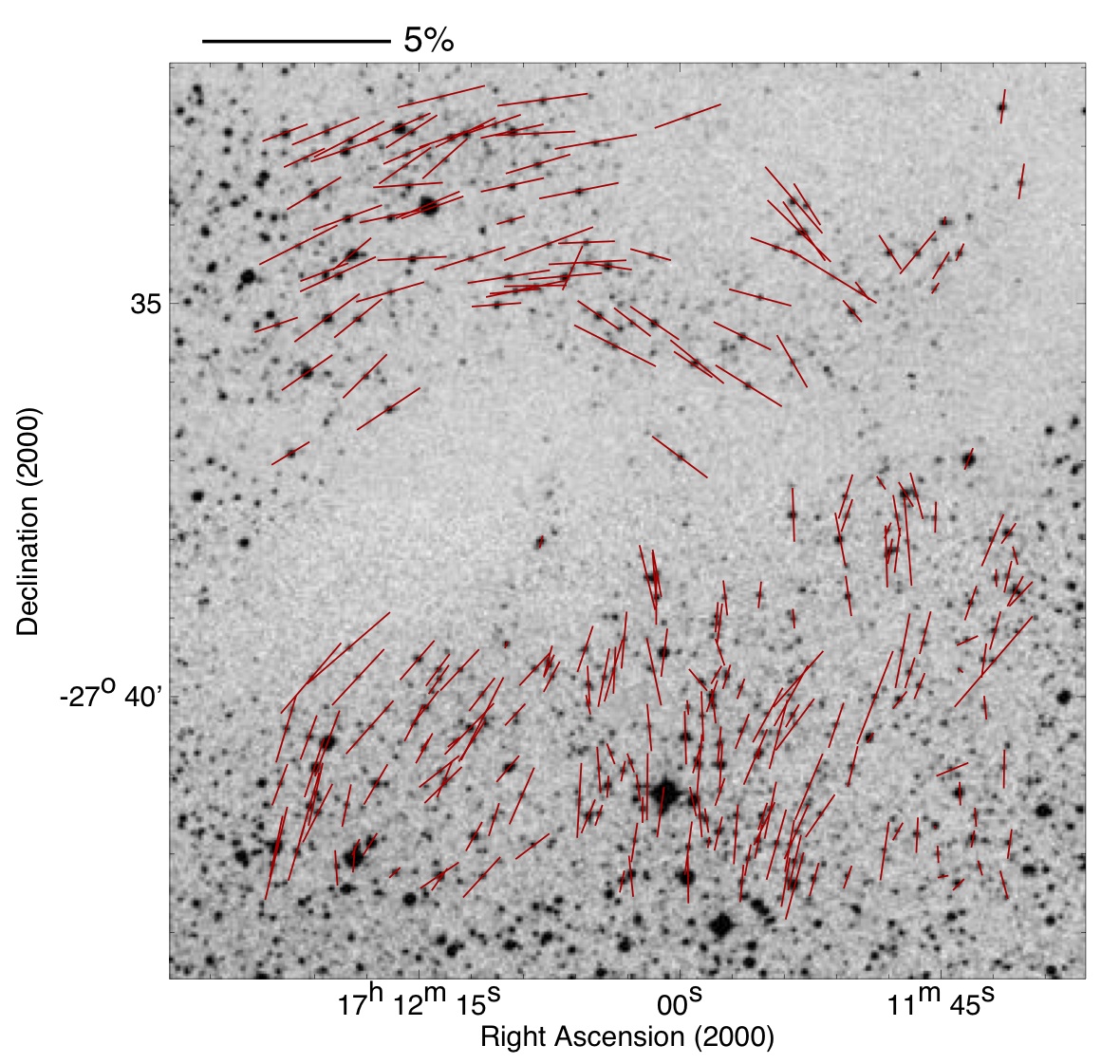}
\caption{Polarization vectors overlaid on the optical image of Field 06.
All measured polarization for this field are represented
in this figure and not only the ones having $P/\sigma_P \ge 5$. The observed
orientation of the polarization vectors seems to embrace the dust core whose 
existence is suggested by the scarcity of observable stars to the left of the center.
The length of the vectors correlates linearly with the degree of
polarization according to the scale indicated over the left-hand corner.}
\label{field_06}
\end{figure}

Although the division between the regions denominated B\,59 and {\it stem} was
chosen rather arbitrarily \citepalias[in][it is characterized by a rising on the 
degree of polarization]{AFG08}, one notice an interesting feature in the 
{\it bottom panel} of Fig.\,\ref{polak}. The polarization efficiency seems to
increase along the dust filaments probed by our sample when we move from
the B\,59 region to the {\it stem}. It happens only for the fields of the {\it stem}
shown in Fig.\,\ref{field_a}, after that, the ratio $p/A_V$ returns to the typical 
value of $\sim$1\,\%\,mag$^{-1}$ observed for B\,59 and the remaining fields in
the {\it stem}. This behaviour can be an indication that distinct physical regimes may 
be acting on different fragments of the {\it stem}. For instance, variations of the 
value of $p/A_V$ may arise where the local magnetic field is not orthogonal to 
or its direction varies along the line-of-sight, or where the processes responsible 
for grain alignment change for some reason (grain 
composition, size, shape, etc). The interested reader will find a good review on
the efficiency of grain alignment in the work by \citet[][and references therein]
{WHL08}. It is worthwhile to mention that the
point where the polarization efficiency returns to its typical value of 
$\sim$1\,\%\,mag$^{-1}$ almost coincides with the place where one noticed the
value of the mean polarization angles started decrease (see 
Fig.\,\ref{meanangle}). 

All these results reinforce once more how interesting is the Pipe nebula and
suggest that this complex may be a testbed for different theories of dust grain 
alignment efficiency.
  
\section{Fields showing interesting polarization distributions}\label{ind_field}

Inspection of Figs.\,\ref{field_a} to \ref{field_d} shows that some of the observed
fields present remarkable polarization geometries. For many of them one clearly
note that the obtained polarization angles for the objects in the field suggest a
multicomponent, or in some cases a hoop-like, distribution. As one have seen, 
in most cases the mean polarization vector is aligned perpendicularly to the long
axis of the Pipe nebula, but there is the case of Field 20 (see 
Fig.\,\ref{meanangle}) where the distribution of polarization angles does not 
follow the average behaviour for the region. Below we introduce three of the 
most interesting observed fields, and comment on the fields having high
mean polarization ($\langle P \rangle \ge 10\%$).

\subsection{Field 06}\label{fld06}

\begin{figure}
\plotone{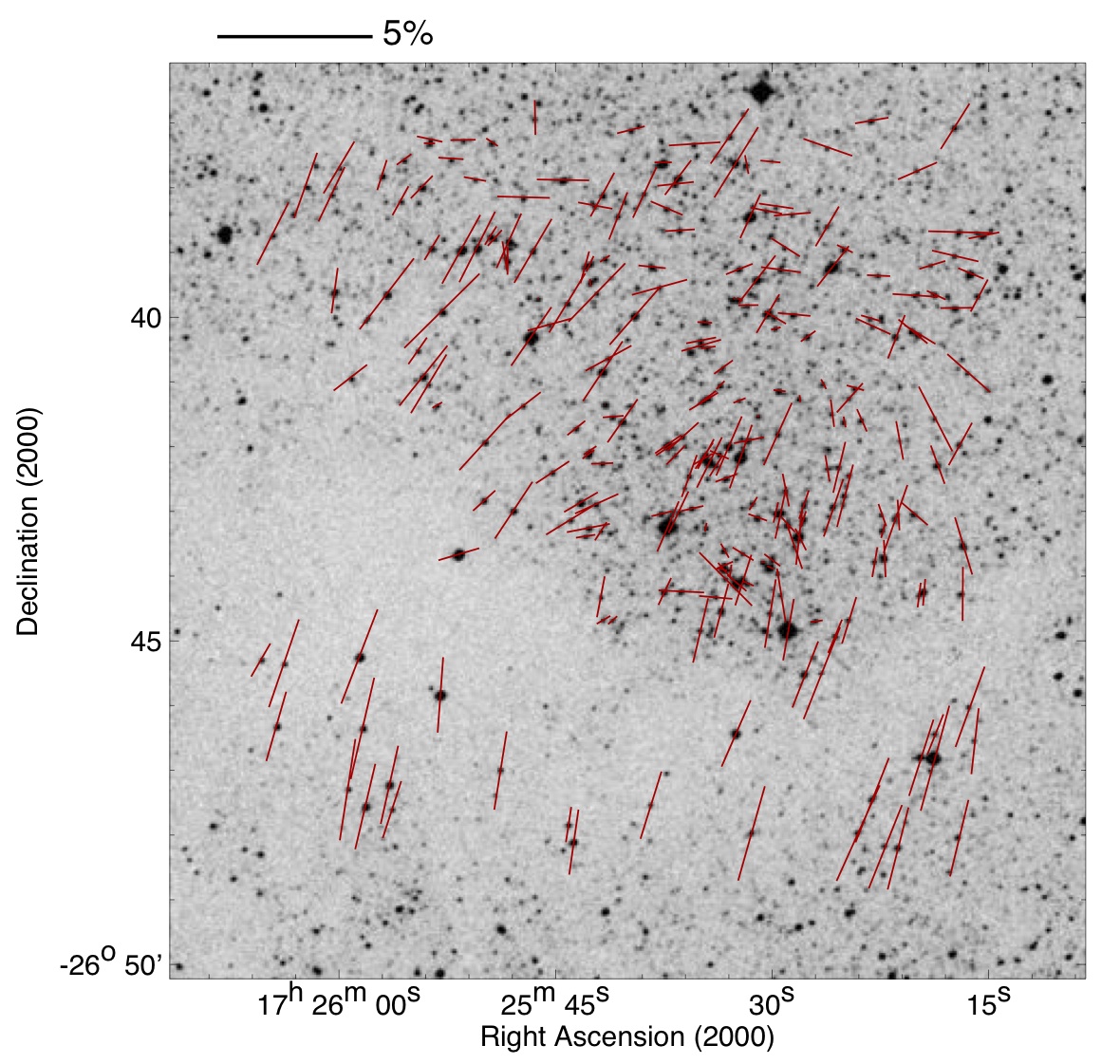}
\caption{Same as Fig.\,\ref{field_06} for Field 26.}
\label{field_26}
\end{figure}

The polarization map for Field 06 is shown in Fig.\,\ref{field_06}, one of the 
most interesting distribution in our survey. This is one of the four fields we have 
identified in Fig.\,\ref{meanangle} as having a mean polarization angle which 
seems to disagree from the pattern observed for the cloud complex. In order to
emphasize the geometry of the magnetic field in this region, all star for which 
polarization has been measured are represented in the map. The polarization 
vectors seem to suggest that the local magnetic field follows the 
border of the dust cloud evidenced by the higher interstellar
absorption noticed to the left of the center. Has this core been modeled by the 
field or, on the contrary, was the field shaped by the core? In any case, this 
seems to be an interesting region which deserves further investigation.

\subsection{Field 26}

The polarization map obtained for Field 26, Fig.\,\ref{field_26}, seems to be 
the result of a mixture of two distributions, a main component centered around 
160\degr\ (see also the histogram introduced in Fig.\,\ref{field_c}), combined 
with an hoop-like component. An interesting point is that, as one may observe 
in the polarization map, the surveyed area seems to show different characteristics
toward directions located northern and southern of the densest parts of 
the cloud --- visually characterized by the absence of stars. Apparently, at the 
south only the main component of the distribution ($\theta \sim 160\degr$) 
is present, while at the north we observe the presence of both distributions. 
An inspection of Fig.\,\ref{field_c} shows that the northern part of this field probes 
a more diffuse part of the interstellar material, as what happens in the case 
of Field 27 (see below), while the southern stars have line-of-sight toward a
volume presenting higher extinction.  
One of the cores studied by \citet{Frau10}, who used the 
IRAM 30-m telescope to carry out a continuum and molecular survey toward 
four of the starless cores from the list of \citet{ALL07}, is Core 48, which is
associated to the higher interstellar absorption shown in Fig.\,\ref{field_26}.  
The radio data indicates that, although being very diffuse, this core has a 
strong dust emission, and their molecular analysis suggests that chemically it
seems to be in a very early stage of evolution.

\subsection{Field 27}

\begin{figure}
\plotone{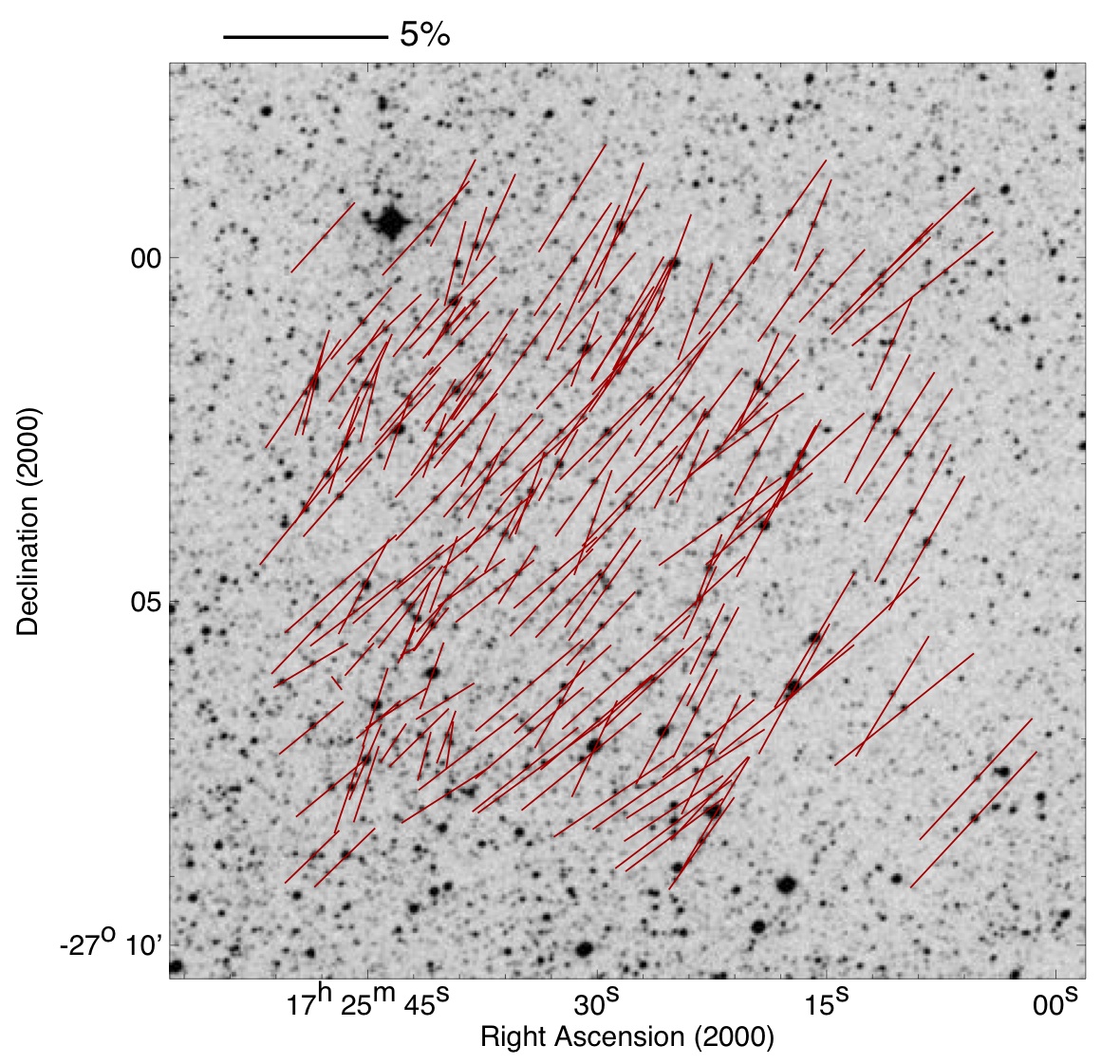}
\caption{Same as Fig.\,\ref{field_06} for Field 27.}
\label{field_27}
\end{figure}

There is no dense core associated to the volume probed by this field, and
it is other of the fields having mean polarization angle not fitting in the main 
pattern of mean position angles, as defined in Fig.\,\ref{meanangle}. The 
distribution of polarization vectors shown in Fig.\,\ref{field_27} (see also the 
histrogram of polarization angles shown  in Fig.\,\ref{field_c}) clearly shows 
a bimodal distribution with mean angles values centered on $\sim$135\degr\ 
and $\sim$155\degr. Both components seem to be well distributed all over 
the surveyed field.

\subsection{Distribution of polarization and position angles as function of the 2MASS $K_S$ magnitude for Fields 26 and 27}

\begin{figure}
\epsscale{1.10}
\plotone{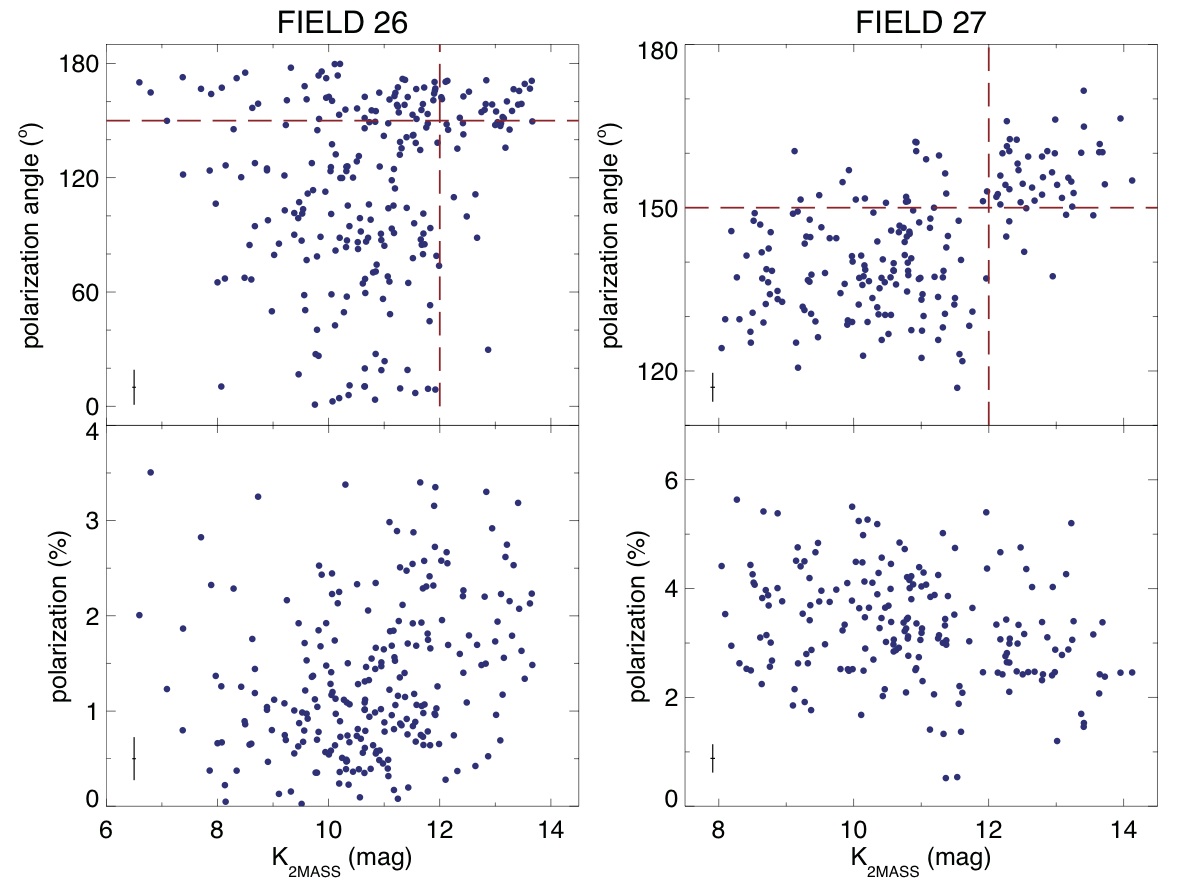}
\caption{Distribution of polarization degrees ({\it bottom panels}) and polarization
angles ({\it top panels}) as a function of the $K$ magnitude. Data obtained for
Field 26 is represented on the {\it left panels} and for Field 27 on the {\it right
panels}. All observed stars in each field were used to construct these diagrams. 
The horizontal and vertical dashed lines, represented in both {\it top panels}, 
were arbitrarily positioned at $\theta=150\degr$ and $K_{\rm 2MASS} =12\fm0$, 
respectively (see text). Mean uncertainties of the 
quantities are indicated by the horizontal and vertical bars on the lower left corner
of each diagram.}
\label{field_26_27}
\end{figure}

The {\it top panels} of Fig.\,\ref{field_26_27} display the measured polarization
angles as function of the 2MASS $K_S$ magnitude for Fields 26 and 27. 
An interesting result comes out from these diagrams. One clearly notice that 
the distribution shown by Field 27 ({\it right panel}) is rather defined by the 
stellar $K_S$ magnitude and occupies different regions of the diagram. Stars 
having $K_S \ga 12^{\rm m}$, that is statistically populated by main sequence
stars, as already mentioned in \S\,\ref{mean_Av}, are mainly associated to 
the component having higher mean angle, while stars having 
$K_S \la 12^{\rm m}$, statistically populated by giant stars, are basically  
associated to the component having lower mean angle. This is indicated by 
the horizontal and vertical dashed lines positioned at $\theta = 150\degr$ 
and $K_S = 12\fm0$, respectively. 

The distribution presented by Field 26 is rather different but shows some of 
the characteristics presented by Field 27. For the sake of comparison,
we have represented the same horizontal and vertical dashed lines in both
diagrams. While the polarization angles observed for Field 27 are restricted
between $\theta \sim 120\degr$ and 170\degr, Field 26 presents basically all
values of polarization angles. However, as observed for Field 27, most of the
stars in Field 26 fainter than $K_S = 12^{\rm m}$ has polarization angle larger
than $\sim$140-150\degr, suggesting that the same kind of interstellar 
structures may be present toward both line-of-sights, which are separated
about 20$^\prime$ from each other. 

It is also interesting to compare the distribution of degree of polarization as a
function of the stellar magnitude ({\it bottom panels}). First of all, one notices that
although Fig.\,\ref{field_c} seems to indicate that the line-of-sight toward 
Field 27 is less affected by interstellar absorption than Field 26, the measured
polarization for the latter is generally smaller than the one obtained for stars
in the former field --- it must be noted, however, that the estimated average
interstellar absorption in \S\,\ref{mean_Av} is essentially the same for both 
fields (see Table\,\ref{table:3}). The $K_S -(J-K_S)$ CMD for the observed stars
in Field 27 suggests that the interstellar absorption toward this line-of-sight is
rather more uniform than the one probed by stars in Field 26, as one should
expect from the dust extinction map obtained by \citet{LAL06} and shown in
detail by our Fig.\,\ref{field_c}. Thus, the estimated average interstellar absorption
for Field 27 is more representative of what we have all over the surveyed 
volume, while the one estimated for Field 26 is a mean between regions showing
rather high absorptions, e.g. toward the southern area of the CCD field, with
regions not so absorbed probed by the stars located in the northern area of the
CCD field.

\subsection{Comments on the Fields with high mean polarization degree}

\begin{figure}
\plotone{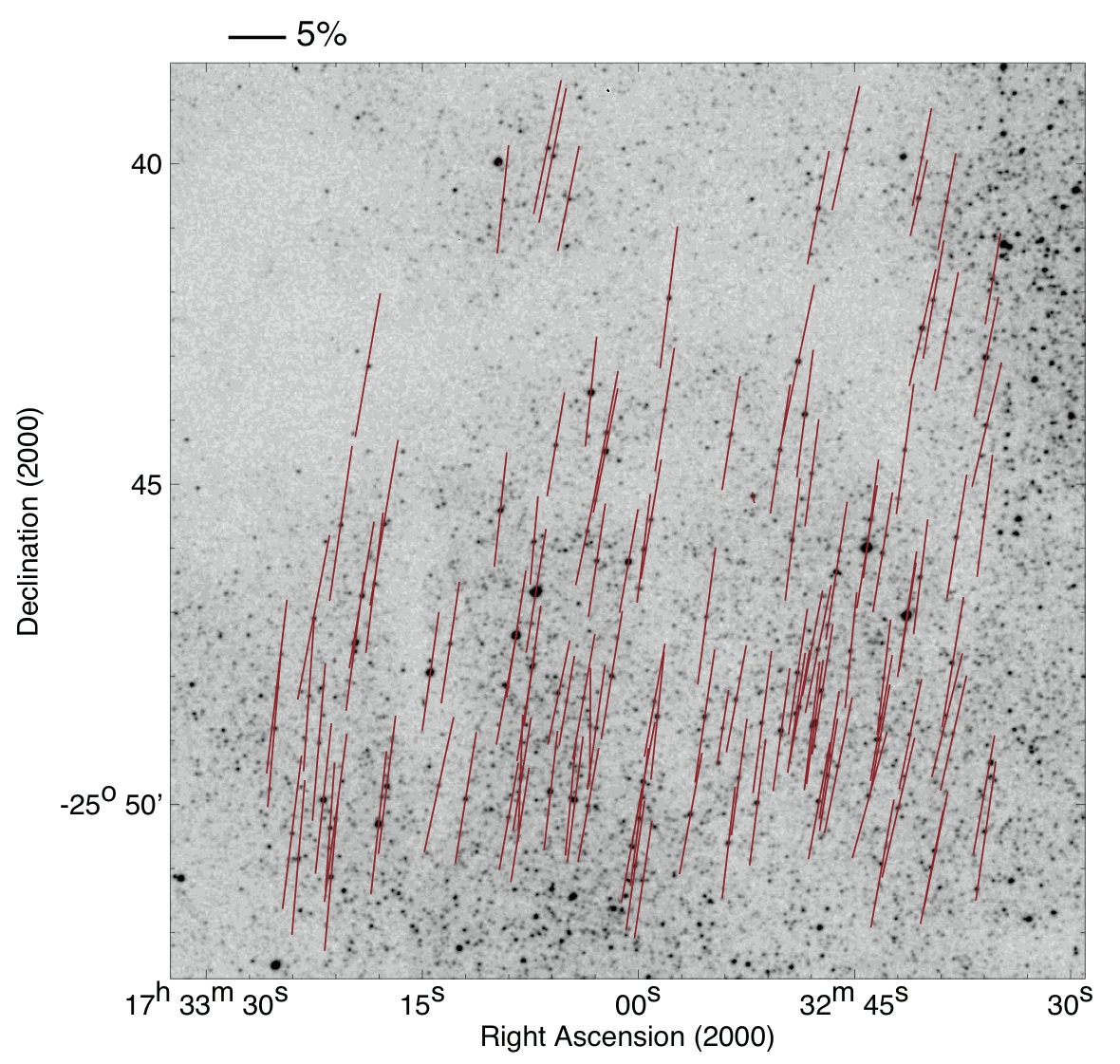}
\caption{Same as Fig.\,\ref{field_06} for Field 35.}
\label{field_35}
\end{figure}

Five of the observed fields present mean degree of polarization $\langle P
\rangle \ge 10\%$, they all lay in the {\it bowl} and are Fields 35, 37, 38, 40, and 
41. In Fig.\,\ref{field_d} these fields are almost aligned along the diagonal crossing 
the image from the upper left-hand to the lower right-hand corner. The main
characteristics of these fields, apart from the high value of observed polarization
degree, is the very low dispersion of polarization angles, which suggests that
the turbulent energy prevailing on the observed cores must be quite low
(see \S\,\ref{structure_function}). In particular, it is noticeable the quite low 
dispersion presented by Field 35 (see also Fig.\,\ref{field_35}), the lowest in 
our survey, with a rather ``normal'' distribution.

\begin{figure}
\epsscale{.85}
\plotone{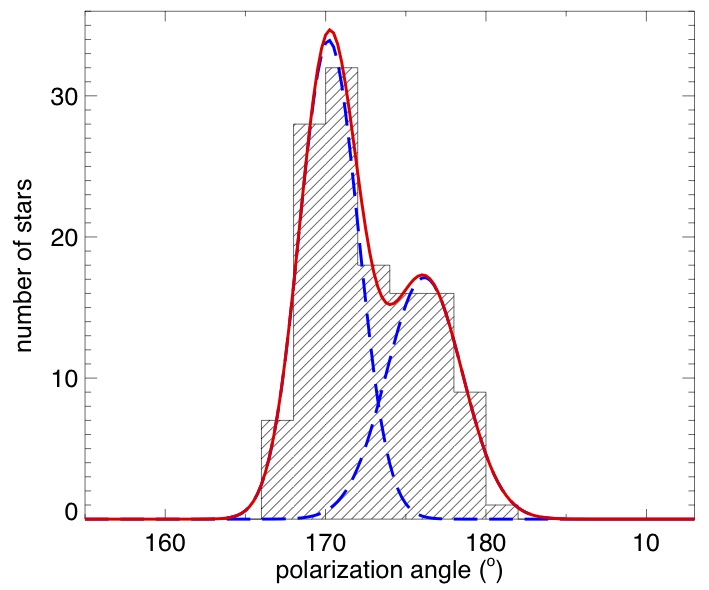}
\caption{The distribution of polarization angles for Field 38 is clearly asymmetric 
suggesting two components. Our best fitting represented by the full line (red) is the 
result of two Gaussian components, one centered at 170\fdg2, $\sigma_{std} 
= 1\fdg83$, and other at 176\fdg2, $\sigma_{std} = 2\fdg36$, represented by 
the dashed lines (blue). All observed stars in this field have $P/\sigma_P > 11$, 
being that most of them have a much larger signal-to-noise ratio, meaning that 
the theoretical uncertainties of the estimated polarization angles are in general
much smaller than 2\fdg6.}
\label{hist_38}
\end{figure}

Although also presenting a very low dispersion, Field 38, 
the one with the highest mean polarization in our survey, shows a fairly 
asymmetry in the observed distribution of polarization angles. As shown in 
Fig.\,\ref{hist_38}, it may be caused by two dust cloud components along the 
observed line-of-sight, each one subject to slightly different orientations of 
ambient magnetic fields. These clouds may be associated to the two main 
velocity components that seem to characterize the kinematics of the  
`bowl'' \citep[e.g.,][]{Mu07}, even though they have not detected two C$^{18}$O
components toward their observed line-of-sight through this field.

The distribution of polarization angles as a function of the 2MASS $K_S$
magnitudes does not present any remarkable feature, unless for the fact that 
the 6 brightest stars in the field ($K_S \la 8\fm0$) have polarization angles 
between 169\degr\ and 172\fdg5, while the remaining stars present a 
rather normal distribution between $\sim 165\degr$ and 180\degr. 

\begin{figure*}
\epsscale{1.10}
\plotone{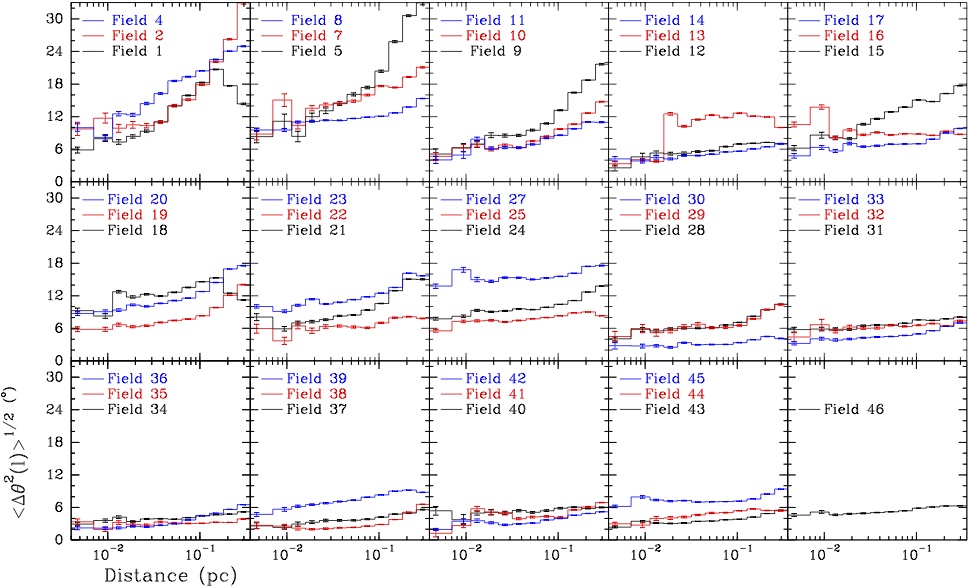}
\caption{Square root of the second order structure function of the polarization 
angles, $\langle \Delta\Phi(l)^2\rangle^{1/2}$, for all the individual fields (except 
for Fields 3, 6, and 26) observed in the Pipe nebula. The units of 
$\langle\Delta\Phi(l)^2\rangle^{1/2}$ are degrees. The upper row panels have a 
larger range of values in the abscissa axis. Fields 1 to 8 are located in B\,59. 
Fields 9 to 24 and 27 are in the {\it stem}. The rest of the fields belong to the 
{\it bowl}. The dashed line shows the $16\fdg6$, which is the value of the 
dispersion of the position angles for turbulent and magnetic equipartition 
\citep{Troland08}.} 
\label{AllFields}
\end{figure*}

Field 40 contains other of the cores observed by \citet{Frau10}, Core 109. 
The radio data show that this object presents a strong dust continuum emission, 
is the densest among the four investigated cores, and one of the most 
massive. The interstellar extinction experienced by the observed stars is very 
nonuniform, ranging from $A_V \approx 2^{\rm m}$ to  $A_V \ga 5^{\rm m}$. 
Interestingly the observed $^{13}$CO molecular emission 
shows a double velocity component (Alves et al., in preparation), which is not 
seen in C$^{18}$O \citep[][ratified by the work in preparation by Alves et 
al.]{Mu07}, and could explain the asymmetry of the distribution of 
polarization angles which, as observed for Field 38, is also noticed for this field
but this time due to a small excess in the  left wing of the distribution
(see distribution introduced in Fig.\,\ref{field_d}). Analyzing the distribution of the 
polarization angles as function of the 2MASS $K_S$ magnitudes one obtained
that this excess is due to stars brighter than $K_S \sim 11\fm5$, which are in
average more affected by the interstellar absorption and present higher mean
degree of polarization. Although located in the {\it bowl}, supposed to be the 
less evolved region of the Pipe nebula, the molecular investigation conducted by 
\citet{Frau10} indicated that the core may be one of the chemically most 
evolved in their molecular survey. 

\section{The Structure Function of the polarization angles in the Pipe nebula}
\label{structure_function}

\subsection{Basic definitions}

The second--order structure function (hereafter $SF$) of the polarization
angles, $\langle \Delta \theta^2(l) \rangle$, is defined as the average of the
squared difference between the polarization angles measured for all pair of
points separated by a distance $l$  (e.g. see equation 5 of Falceta-Gon{\c
c}alves et al. 2008). Thus, the $SF$ give information on the behavior of the
dispersion of the polarization angles as a function of the length scale in
molecular clouds. Recently, it has been used as a powerful statistical tool to
infer information of the relationship between the large-scale and the turbulent
components of the magnetic field in molecular clouds \citep{Falceta08, HKD09,
Houde09}.  Given the large statistical sample of the polarization data in the
Pipe nebula, it is interesting to compute the $SF$ along the Pipe nebula to
scales up to few parsecs. For a qualitatively discussion we will first use the
square root of $SF$, also called the angular dispersion function or $ADF$
\citep{Poidevin10}.  The use of the $ADF$ instead of the $SF$ allows a more
straightforward comparison of the behavior of position angle dispersion as a
function of the length scale.  Then, we use the $SF$ to compare our statistical
sample with the previous works \citep{Falceta08, Houde09}.

\subsection{Qualitative analysis}

\begin{figure*}
\epsscale{.80}
\plotone{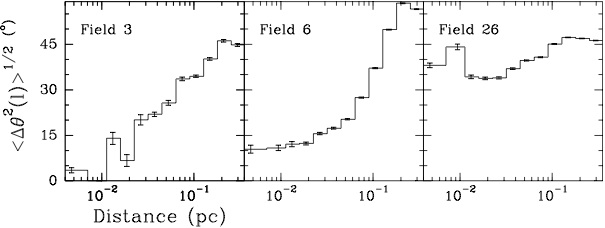}
\caption{Same as Fig.\,\ref{AllFields} but for Fields 3, 6 and 26, which show a 
higher dispersion.} 
\label{Field26}
\end{figure*}

\begin{figure}
\epsscale{1.1}
\plotone{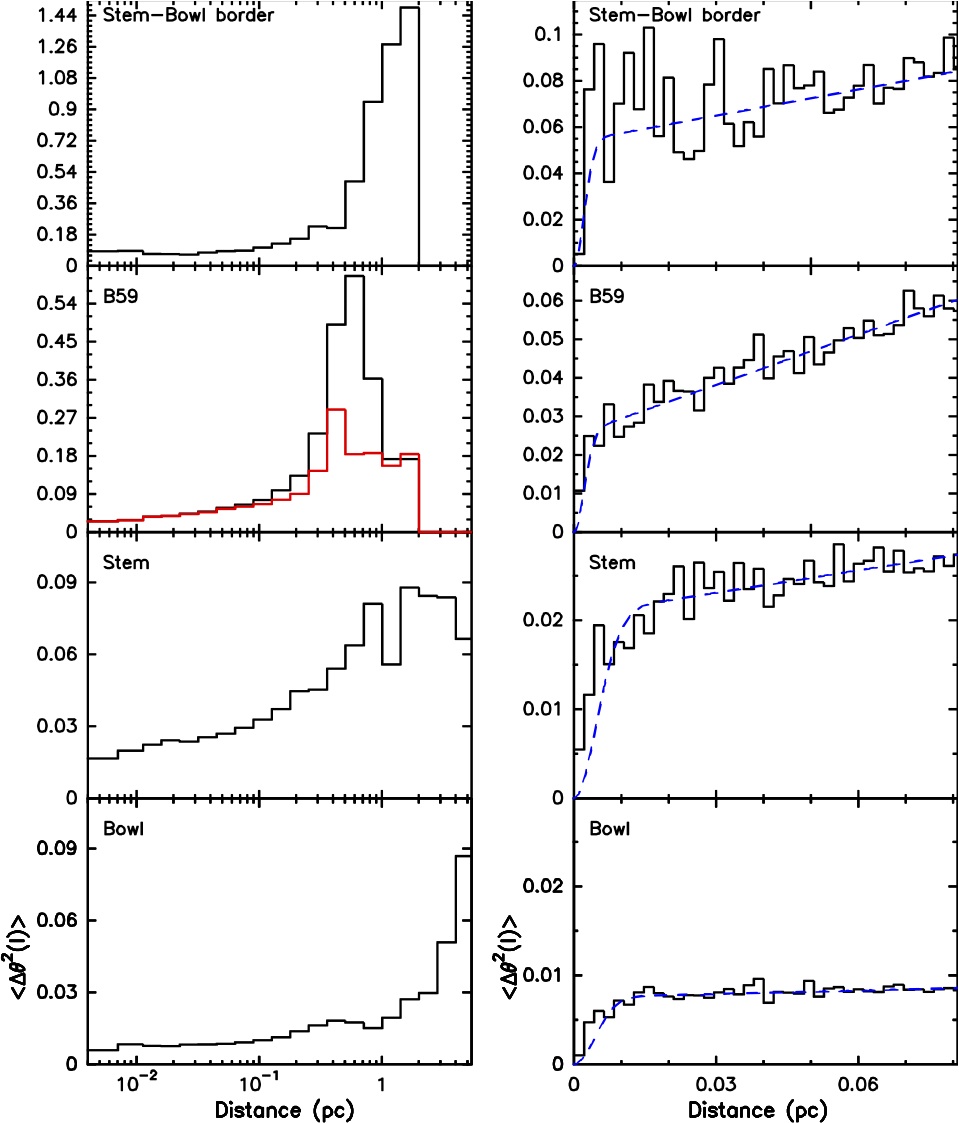}
\caption{Second--order structure function of the polarization angles, 
$\langle\Delta\Phi(l)^2\rangle$, for the five distinctive regions in the Pipe nebula 
(the list of fields associated with each region is give in Table\,\ref{table:4}). 
Right and left panels show the smallest and largest scales for each region, 
respectively.  The gray (red) histogram for B\,59 shows the structure function 
without Fields 3 and 6. The dashed-lines (blue) indicates the best fit of equation 
\ref{EqHoude} for distance up to 0.07 pc.
} 
\label{StFnPipa}
\end{figure}

We have first computed the $ADF$ for all the individual fields using a
logarithmic scale between $8''$ (5.6~mpc) and $11\farcm8$ (0.35~pc). This range
was selected in order to have a good statistical sample. Figure\,\ref{AllFields}
shows the $ADF$ for all the Fields but 3, 6, 26. These three fields, which are
the ones that exhibit the highest polarization angle dispersion (see
Table~\ref{table:3}), are shown separately in Fig.\,\ref{Field26}.  There is a clear trend
in the distribution of the $ADF$ along the Pipe nebula (see
Fig.~\ref{AllFields}). On one hand, fields in B\,59 (1--8) not only show a higher
polarization angle dispersion at all the observed scales but the $ADF$ slope is
the highest. A steep slope is an indication that the large-scale magnetic field
orientation in the plane of the sky changes significantly. Fields 3 and 6 have
dispersion values at scales larger than $\simeq 0.1$~pc close to the expected
maximum dispersion that would be obtained in case of a  purely random
polarization angle distribution, $\simeq 52\arcdeg$, \citep{Poidevin10}. As
pointed in \S\,\ref{fld06} for Field 6 this is due to a strongly distorted field
surrounding a core.  On the other hand, all the Fields in the {\it bowl} (28--46) not
only have a remarkably small dispersion of the position angles \citep{AFG08} but
this trend is also observed in the $ADF$  at all the observed scales. Indeed and
in contrast with B\,59, the almost flat slope of the $ADF$ in the {\it bowl} Fields
indicate that the projected magnetic field in the plane of the sky is very
uniform. The $ADF$ behavior of the {\it stem} Fields is intermediate between that of
B\,59 and the {\it bowl}.  However, the global $ADF$ properties of field 26 differ from
the general trend found in the {\it stem}: it shows an unusual high dispersion,
$\simeq 40\arcdeg$ at all scales. Compared with other Fields with also a high
dispersion  (e.g. 3 and 6) the $ADF$ slope of Field 26 is
relatively flat. Field 27, the one with a bimodal distribution (see \S\,5.3),
shows a similar $ADF$ behavior to Field 26 but a smaller level: $ADF \simeq 17\arcdeg$ 
at all scales. Because of the peculiarities of these two Fields, we 
treat them as a distinctive region in the Pipe nebula for the $SF$ analysis. 
Here on, we call this region as the ``{\it stem--bowl} border''. We also include 
Field 20 in this region because its average mean direction is quite different 
from the ones of the rest of the {\it stem} (see Fig.\,\ref{meanangle}) and it is relatively 
near to Fields 26 and 27.

\begin{figure}
\epsscale{0.85}
\plotone{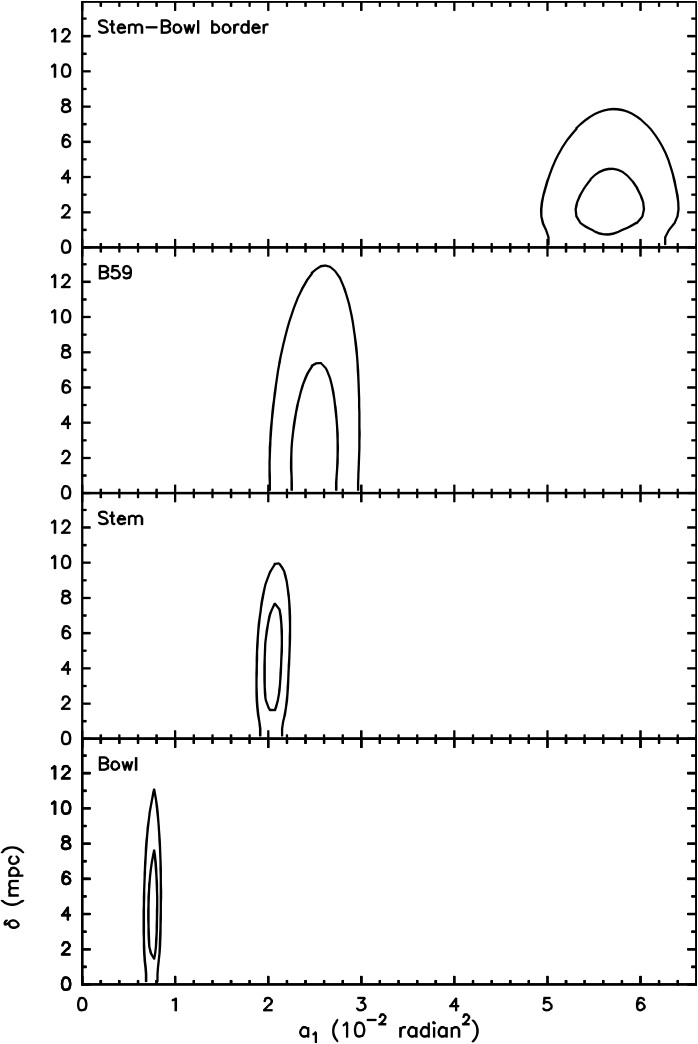}
\caption{Plot of the set of the solutions for the $\delta_t$ and $a_1$ parameters of the $SF$. 
The inner and outer contours show the 63.3\% and 99\% confidence regions of the $\chi^2$, 
respectively.} 
\label{ChiDos}
\end{figure}

Figure\,\ref{StFnPipa} shows the $SF$ for the  four distinctive regions within
the Pipe nebula: B\,59, the {\it stem} (except Fields 20, 26 and 27), the {\it
bowl} and the {\it stem--bowl} border. By combining the different fields of each
region, the $SF$ can be computed at larger scales, up to few parsecs (see left
panels of Fig.\,\ref{StFnPipa}). The general trend described in the previous
paragraph for the $ADF$ also applies for the $SF$. For example, it is 
remarkable that the {\it bowl} shows very low $SF$ values up to scales of few 
parsecs, with position angle dispersion lower than $\simeq 10\arcdeg$. 
Yet, B\,59 and the ``{\it stem--bowl} border''  show an abrupt increase of the 
$SF$ at scales larger than 0.3--0.5~pc. For B\,59 this is due to the high 
PA dispersion Fields 3 and 6.  Indeed, if these two fields are excluded, the 
resulting $SF$ is smoother at these scales.

\subsection{Comparison with \cite{Houde09}}

\begin{deluxetable*}{cccrcc}
\tablecaption{Structure function parameters for the Pipe nebula\label{table:4}}       
\tablewidth{0pt}
\tablehead{
\multicolumn{1}{c}{Region} &
\multicolumn{1}{c}{$a_0$} &
\multicolumn{1}{c}{$a_1$} &
\multicolumn{1}{c}{$\delta_t$} &
\multicolumn{1}{c}{($\delta B^2_t/B^2_0$)\tablenotemark{a} } &
\multicolumn{1}{c}{Fields} 
\\
\multicolumn{1}{c}{} &
\multicolumn{1}{c}{($\frac{\rm radian^2}{\rm pc}$)} &
\multicolumn{1}{r}{(radian$^2$)} &
\multicolumn{1}{r}{(mpc)} &
\multicolumn{1}{r}{} &
\multicolumn{1}{c}{} 
}
\startdata
B\,59		& 0.44	&  0.025	& 2.1	&0.4& 1--8 \\
{\it stem} 		& 0.08	&  0.021	& 4.8	&0.2& 9--19,21--25\\
{\it stem--bowl}	& 0.38	&  0.054	&$\le2.1$	&0.8& 20,26,27 \\
{\it bowl} 		& 0.01	&  0.008	& 4.4	&0.1& 28--46 
\enddata
\tablenotetext{a}{Estimated for $N=30$ (see text).}
\end{deluxetable*}

The larger statistics obtained by dividing the Pipe in four region also allows
to better sample the smaller scales. The right panels of Fig.\,\ref{StFnPipa}
show that at scales of few hundredths of a parsec the $SF$ increases linearly
with length scale. Yet, at scales of $\la0.01$~pc, the $SF$ drops fast when
approaching to the zero length scale. This is a clear indication that we are
starting to resolve the correlation length scale for the turbulent magnetic 
field component, which is the scale at which the turbulent energy is dissipated. 
In order to approximately estimate  the turbulent length scale we follow the 
recipe given in the detailed analysis carried out by \citet{Houde09}, where 
they assumed a Gaussian form for the autocorrelation function for the 
turbulence. We use equation\,44 from \citet{Houde09} taking into account that the 
effective angular resolution of the optical polarization data can be considered 
zero. Therefore, in that equation the angular resolution term is $W=0$. 
Thus, equation\,44 from \citet{Houde09} can be rewritten as:

\begin{equation}\label{EqHoude}
\langle \Delta \Phi (l)^2  \rangle  \simeq 
a_0 \,l + a_1 [ 1 - e^{-l^2/2 \delta_t^2}]  
\end{equation}
The first term, $a_0 \, l$, gives the large-scale magnetic contribution to the
$SF$ (note that we have adopted a linear dependence instead of the original 
$l^2$ dependence of the aforementioned Eq.\,44).  The second term 
corresponds to the turbulent contribution to the $SF$. $\delta_t$ is the 
turbulent length scale and $a_1$ is a function of the large--scale magnetic field 
strength, $B_0$, the turbulent component of the magnetic field, $\delta B_t$,
and of $N$, the number of turbulent correlation lengths along the line of 
sight\footnote{Note that this term is equivalent to number 
of the magnetic field correlation length along the line of sight introduced by \cite{MG91}}:
\begin{equation}\label{EqHoude2}
a_1 =  (2/N) \, (\delta B_t^2/B_0^2)
\end{equation}
$N$ can also be understood as the number of independent turbulent cells 
along the line of sight \citep{Houde09}. We have used equation\,\ref{EqHoude} 
to fit the $SF$ data in the four Pipe regions 
for the scale range shown in the right panels of Fig.\,\ref{StFnPipa}. These are the 
scales in which the large-scale magnetic field contribution to the $SF$ is basically 
linear. For each of the four Pipe regions a $\chi^2$ analysis was carried out 
to find the best set of solutions for the free parameters $a_0$, $a_1$ and 
$\delta_t$.  The best-fit solutions obtained are given in Table\,\ref{table:4} and 
the 99\% and 67\% confidence intervals for $a_1$ and $\delta_t$ are shown in 
Fig.\,\ref{ChiDos}. The  dashed blue line in the right panels of  
Fig.\,\ref{StFnPipa} show the best solution for each region. We find that the 
turbulent  correlation length,  $\delta_t$, is in all cases of few milliparsecs. Given that 
the assumption of the Gaussian form for the turbulence autocorrelation function is 
not correct, the found values should be taken as an approximation. In any case, 
the right panels of Fig.~\ref{StFnPipa} show that the turbulent  correlation length 
should be $\la0.01$\,pc. Indeed, at the 99\% confidence level the 
$\chi^2$ analysis provides an  upper limit for  $\delta_t$  of $\simeq 12$\,mpc 
(see Fig.~\ref{ChiDos}).  This upper limit is slightly lower that the $\delta_t$ 
found  for OMC-1, 16\,mpc, from submm polarization observations 
\citep{Houde09}. 

From Eq.\,\ref{EqHoude2}  we can estimate the turbulent to magnetic 
energy ratio, $\delta B_t^2/B^2_0$, if the number of independent 
turbulent cells, $N$, is known.  For optical polarization data, the number 
of independent turbulent cells is $N \simeq \Delta/(\sqrt{2 \pi} \, \delta_t)$
\citep{Houde09}, where $\Delta$ is the cloud thickness.   
$\Delta \simeq N({\rm H}_2) / n({\rm H_2})$, where $N({\rm H}_2)$ and 
$n({\rm H_2})$ are the column and volume densities of the molecular 
gas traced by the optical polarization data. $N({\rm H}_2)$ can be obtained 
from the typical visual extinction of the observed fields.
The average visual extinction for the  {\it bowl} is 3.6\,mag, whereas
for the rest of the regions is 2.1\,mag. Using the standard conversion to column
density \citep{Wagenblast89}, these values yields to $N({\rm H}_2) \simeq
4.5\times 10^{21}$ and $2.6\times 10^{21}$~cm$^{-2}$ for the {\it bowl} and for
the rest of the regions, respectively. For the observed fields, \citetalias{AFG08} 
estimated that the volume density of the  gas associated with the optical 
polarization is $n({\rm H_2}) \simeq 3 \times 10^3$~cm$^{-3}$. Therefore, 
the cloud thickness of 0.5\,pc for the {\it bowl} and of 0.3\,pc for the rest 
of the regions. With these values and using for $\delta_t$ the range at the 
67\% confidence interval  (see Fig.~\ref{ChiDos})  we obtain that $N$ 
ranges between 25 and 100 for the {\it bowl} and 15 to 60 for the rest of the 
regions.  Nevertheless,  a high value of $N$ will also reduce significantly 
the observed  polarization level.  But all the {\it bowl} fields and many of 
the {\it stem} fields have polarization levels of  4--15\% and 3--4\%, 
respectively.   Therefore, it is unlikely the case of a high $N$, at least, for 
these two regions. Indeed, \citet{MG91} estimated that for optical 
polarization observations  $N$ is expected to not be larger than 
$\simeq 14$. \citet{Houde09} found $N\simeq21$ for OMC-1 from submm 
dust polarization observations that trace significantly larger column 
densities. Therefore, we tentatively adopt a relatively high value of $N=30$. 
For this case, the magnetic field appears to be  energetically dominant 
with respect to turbulence in the Pipe nebula except for the ``{\it stem--bowl} 
border'',  where magnetic and turbulence energy appear to be in 
equipartition (see Table~\ref{table:4}).

\subsection{Comparison with \cite{Falceta08}}

\citet{Falceta08} carried out simulation of turbulent and magnetized molecular 
clouds computing the effect on the dust polarization vectors in the 
plane-of-the-sky  for cases with super-Alfv\'enic and sub-Alfv\'enic turbulence 
(i.e., clouds energetically dominated by turbulence and magnetic fields, 
respectively).  They computed the $SF$ derived from dust polarized 
emission as well as from optical polarization using background stars for the 
different sub and super-Alfv\'enic cases, and for different angles of the 
magnetic  field with respect to the line of sight  (see Figs. 6 and 11 of this 
paper). The SF for super-Alfv\' enic turbulence is clearly higher than the one for 
sub-Alfv\'enic turbulence: The $SF$ ranges from 0.4  at the smallest 
scales up to $\simeq 1.0$ to the highest scales (see central panel of Fig. 6 
from Falceta-Gon\c{c}alves et al. 2008). For the case of sub-Alfv\'enic 
turbulence such a high values of the $SF$ are reached only in the cases where 
the magnetic field direction is close to the line of sight. For the other cases
of sub-Alfv\'enic turbulence, $SF \la 0.5$. Comparing the $SF$ 
obtained in the four Pipe nebula regions (Fig.~\ref{StFnPipa}) with the
results of  \citet{Falceta08} it is clear that B\,59, the {\it stem}, and the 
{\it bowl} are compatible with the presence of sub-Alfv\'enic turbulence. The 
behavior of the $SF$ for the {\it stem--bowl} border ($SF$ from 0.1 at the 
smallest scale to $\ga 1.0$ at the larger scales) may indicate the case of
sub-Alfv\'enic turbulence with a magnetic field near the line of sight rather
than  super-Alfv\' enic  turbulence.  Indeed, the only individual field in the 
Pipe nebula that at all scales have a $SF$ compatible with the 
super-Alfv\' enic turbulence is Field 26.

\subsection{Summary of the $SF$ analysis}

The comparison of the $SF$ derived from our optical polarization data with the ones 
derived in the works by \citet{Houde09} and \citet{Falceta08}, indicated that the 
Pipe nebula is a magnetically dominated molecular cloud complex and that the
turbulence appears to be sub-Alfv\'enic. Only the region we call the
{\it stem--bowl} border, in particular Field 26, appears to have
a behavior that is compatible with super-Alfv\' enic turbulence.
A similar situation seems to apply to the well investigated low mass star 
forming region in the Taurus complex where there is evidence for a molecular
gas substrate with sub-Alfv\'enic turbulence and magnetically subcritical
\citep{HGO08, Nakamura08}.  \citet{HF07} also found that in Taurus, the magnetic 
fields are dynamical important, although they found that they are trans-Alfv\'enic.  
In addition, analyzing the polarization angles at different scales using optical 
and submm observations in several molecular cloud yield \citet{Li09} to
suggest that these clouds are also sub-Alfv\'enic.

\section{Summary}

The Pipe nebula has proved to be an interesting interstellar complex where to
investigate the physical processes that forestall the stellar formation phases. 
The polarimetric survey analyzed in this work covers a small fraction only of
the entire Pipe nebula complex, and there is no doubt that new data is highly
desired in order to verify some of the speculations settled in this investigation.
In \citetalias{AFG08}, we suggested that the Pipe 
nebula, a conglomerate of filamentary clouds and dense cores, is possibly experiencing different stages of evolution. From the point of view of the
global polarimetric data alone, we proposed three evolutionary phases from 
B\,59, the most evolved region, to the {\it bowl}, the youngest one, however, 
the real scenario seems to be much more complicated than that. As 
demonstrated by \citet{Frau10}, from the point of view of the chemical 
properties derived for four studied starless cores, there does not seem to be 
a clear correlation between the chemical evolutionary stage of the cores and 
their position in the cloud.

In addiction, the polarimetric analysis conducted here suggests that,

\begin{enumerate}
\item Although the unusually high degree of polarization, observed for numerous 
stars in our sample, the probed interstellar dust does not seem to present any 
peculiarity as compared to the common diffuse interstellar medium. In fact, 
the fields where the high polarization were observed show a polarization 
efficiency of the order of $p/A_{\rm V} \approx 3$\,\%\,mag$^{-1}$, which is 
the typical maximum value universally observed for the diffuse interstellar 
medium.
\item Basically all observed fields in B\,59 and the Pipe's {\it stem} present 
an estimated polarization efficiency of the order of 
$p/A_{\rm V} \approx 1$\,\%\,mag$^{-1}$, and all so far known candidate 
YSOs presumed associated to the Pipe nebula were found in those regions.  
\item  While the value of the mean polarization angle obtained for fields 
toward volumes not associated to the densest parts of the 
main body of the Pipe nebula seems to remain almost constant, the same does 
not happens for fields presenting large interstellar 
absorption, suggesting that the uniform component of the magnetic field 
permeating the densest filaments of the Pipe nebula shows systematic 
variations along the main axis of the dark cloud complex. 
\item Analysis of the second--order structure function of the polarization angles 
suggests that in the Pipe nebula the large scale magnetic field dominates 
energetically with respect to the turbulence, i.e. the turbulence is 
sub-Alfv\'enic.  Only in a localized region between the {\it bowl} and the  {\it stem}
turbulence appear to be dynamically more important. 
\end{enumerate}
   
\acknowledgements

We thank the staff of the Observat\'orio do Pico
dos Dias (LNA/MCT, Brazil) for their hospitality and invaluable help during
our observing runs. We made extensive use of NASA's Astrophysics Data System 
(NASA/ADS) and the SIMBAD database, operated at CDS, Strasbourg, France. 
The optical images used to prepare Figs.\,\ref{field_06} to \ref{field_27} and
\ref{field_35} were retrieved from the 2nd Digitized Sky Survey (DSS2) produced 
at the Space Telescope Science Institute under US Government grant 
NAG W-2166. This research has made use of the NASA/IPAC Infrared Science 
Archive, which is operated by the Jet Propulsion Laboratory, California Institute of 
Technology, under contract with the National Aeronautics and Space 
Administration. We are grateful to Drs. A. M. Magalh\~aes and A. Pereyra 
for providing the polarimetric unit and the software used for data reductions,
and to Dr. M. Lombardi for generously providing us the Pipe nebula IR extinction
map used to prepare Figs.\,\ref{pipe_areas}, and from \ref{pol_ang}
to \ref{field_d}.  This research  has been partially supported by CEX 
APQ-1130-5.01/07 (FAPEMIG, Brazil), CNPq (Brazil), and  AYA2008--06189--C03--02 (Ministerio 
de Ciencia e Innovaci\'on, Spain).

\end{document}